\documentclass[]{rsos}
\usepackage{hyperref,aas_macros}

\begin{document}

\title[The transiting dust clumps of RZ~Psc]{The transiting dust clumps in the evolved
  disk of the Sun-like UXor RZ~Psc}

\author[Grant M. Kennedy et al]{Grant M. Kennedy$^1$,
  Matthew A. Kenworthy$^2$,
  Joshua Pepper$^3$,
  Joseph E. Rodriguez$^{4,5}$, 
  Robert J. Siverd$^6$,
  Keivan G. Stassun$^{5,7}$, \&
  Mark C. Wyatt$^1$}

\address{\footnotesize{$^1$ Institute of Astronomy, University of Cambridge, Madingley Road, Cambridge CB3
  0HA, UK \\
  $^2$ Leiden Observatory, Leiden University, PO Box 9513, NL-2300 RA Leiden, the
  Netherlands \\
  $^3$ Department of Physics, Lehigh University, 16 Memorial Drive East, Bethlehem, PA
  18015, USA \\
  $^4$ Harvard-Smithsonian Center for Astrophysics, 60 Garden Street, MS-78, Cambridge, MA
  02138, USA \\
  $^5$ Department of Physics and Astronomy, Vanderbilt University, 6301 Stevenson Center,
  Nashville, TN 37235, USA \\
  $^6$ Las Cumbres Observatory Global Telescope Network, 6740 Cortona Dr., Suite 102, Santa
  Barbara, CA 93117, USA \\
  $^7$ Department of Physics, Fisk University, 1000 17th Avenue North, Nashville, TN
  37208, USA \\
}}

\subject{astrophysics, extrasolar planets, stars}


\corres{Grant M. Kennedy \\
\email{gkennedy@ast.cam.ac.uk}}

\begin{fmtext}
\end{fmtext}

\begin{abstract}
  RZ~Psc is a young Sun-like star, long associated with the UXor class of variable stars,
  which is partially or wholly dimmed by dust clumps several times each year. The system
  has a bright and variable infrared excess, which has been interpreted as evidence that
  the dimming events are the passage of asteroidal fragments in front of the host
  star. Here, we present a decade of optical photometry of RZ~Psc and take a critical
  look at the asteroid belt interpretation. We show that the distribution of light curve
  gradients is non-uniform for deep events, which we interpret as possible evidence for
  an asteroidal fragment-like clump structure. However, the clumps are very likely seen
  above a high optical depth mid-plane, so the disk's bulk clumpiness is not
  revealed. While circumstantial evidence suggests an asteroid belt is more plausible
  than a gas-rich transition disk, the evolutionary status remains uncertain. We suggest
  that the rarity of Sun-like stars showing disk-related variability may arise because i)
  any accretion streams are transparent, and/or ii) turbulence above the inner rim is
  normally shadowed by a flared outer disk.
\end{abstract}

\maketitle

\section{Introduction}\label{s:intro}

Disks of gas and dust surround essentially all young analogues of our Sun
(e.g. \cite{2001ApJ...553L.153H}). The lifetime of the gas in these disks is very short
compared to the stellar lifetime, and within a few million years has accreted onto the
star, been lost to space in photoevaporative flows, and contributed to building
planets. The evolution of the dust during this phase is uncertain, but the existence of
gas giant planets makes it clear that planetary building blocks, and of course some
planets, form on a similar or shorter timescale.

Beyond the first few million years a typical star hosts a planetary system, the
components being the planets themselves and a residual disk of small bodies. These
``planetesimals'' -- the asteroids and comets -- make up the ``debris disk'', where the
standard picture is that destructive collisions between them generate a size distribution
of fragments that extends down to micron-sized dust
(e.g. \cite{1993prpl.conf.1253B,2002MNRAS.334..589W,2010RAA....10..383K}).

The state of planetary systems as they emerge from the gas-rich phase is
uncertain. Planets' locations are not finalised at this epoch, but may move by
interacting with other stars, planets, and/or planetesimals in the system
(e.g. \cite{2007ApJ...669.1298F,1996Sci...274..954R,2005Natur.435..459T}). Similarly, the
state and origin of the debris disk is uncertain. At stellocentric distances near 1au,
the region of interest in this article, it could be that dust observed at this time is
related to the final stages of planet formation
(e.g. \cite{2010ApJ...717L..57M,2012MNRAS.425..657J}), originates in young analogues of
our Asteroid belt \cite{2013MNRAS.433.2334K}, is a signature of comets scattered inwards
from more distant regions (e.g. \cite{2009MNRAS.399..385B}), or is simply a remnant of
the gas-rich disk that has yet to be dispersed \cite{2014MNRAS.438.3299K}. In the absence
of gas detections that argue for the latter scenario, discerning among these various
scenarios, which are not mutually exclusive, is difficult.

A promising way to probe these inner regions is by observing temporal variability
(e.g. \cite{2014Sci...345.1032M}). Optical and IR stellar variation has been studied for
decades (e.g. \cite{1945ApJ...102..168J,1994AJ....108.1906H}), and has recently been
reinvigorated by large scale efforts
(e.g. \cite{2011ApJ...733...50M,2014AJ....147...82C}) and as a side-effect of large-scale
surveys for transiting planets \cite{2012AJ....143...72M,2013AJ....146..112R}. Of many
different classes of variables, the ones of most interest and relevance here are the
``UXors'', named for the prototypical system UX~Orionis \cite{1994AJ....108.1906H}. These
are usually Herbig Ae and late-type Herbig Be stars \cite{1984A&AS...55..109F}, and
typically show several magnitudes of extinction that is generally attributed to variable
obscuration by circumstellar dust
\cite{1994AJ....108.1906H,1999AJ....118.1043H,2000A&A...364..633N}.\footnote{Not all
  stars occulted by dust are UXors. Two other classes are: those occulted by
  circumstellar material beyond $\gtrsim$10au, such as AA~Tau and V409~Tau which show
  $\gtrsim$year-long dimming events \cite{2013A&A...557A..77B,2015AJ....150...32R}, and
  systems such as $\epsilon$~Aur, J1407, EE~Cep, and OGLE-LMC-ECL-11893 where the
  occultations are attributed to circumsecondary disks
  (e.g. \cite{1999MNRAS.303..521M,2012AJ....143...72M,2014ApJ...788...41D,2015ApJS..220...14K}).}
Three related arguments that favour circumstellar dust as the cause are i) a maximum
depth of dimming events of roughly 3 magnitudes, suggesting that a few percent of the
visible flux is not directly from the star, but scattered off a disk that surrounds the
star and remains visible even when the star itself is completely occulted, ii)
``blueing'', where the star is reddened for small ($\lesssim$1mag) levels of dimming but
returns to the stellar colour (i.e. becomes ``bluer'') for the very deep ($\gtrsim$1mag)
events where the star is mostly occulted -- the reddening indicates dimming by
circumstellar dust, and a stellar colour is typical of light scattered off circumstellar
dust \cite{1988SvAL...14...27G}, and iii) increased polarisation fraction during dimming
events, caused by a greater fraction of the flux being contributed by dust-scattered
light (e.g. \cite{2001ARep...45...51R}, which also shows that the surrounding dust does
not reside in a spherical shell). UXors therefore reveal information on the degree of
non-axisymmetry, the ``clumpiness'', of dust orbiting a star on spatial scales similar to
the star itself. The observations can span multiple orbits to test for repeated dimming
events (e.g. \cite{1999A&A...349..619B}), and by using different bandpasses and
polarisation can estimate dust grain sizes
(e.g. \cite{1988SvAL...14...27G,1994ASPC...62...63G}).

In the majority of UXor-like cases (i.e. those related to obscuration by dust), including
other classes such as ``dippers'' (e.g. \cite{2014AJ....147...82C,2016ApJ...816...69A}),
the processes causing young stars to vary are attributed to gas-rich protoplanetary
disks. For Herbig Ae/Be stars the obscuration is thought to be caused by hydrodynamic
turbulence that lifts dust above the puffed up inner rim of a self-shadowed disk
\cite{2003ApJ...594L..47D}. For the dippers, which are observed around low-mass stars,
the obscuration is attributed to dust in accretion streams that link the inner disk and
the stellar surface, and/or to variations in the height of the inner disk edge
\cite{1999A&A...349..619B,2016ApJ...816...69A,2016arXiv160503985B}. The common theme is
therefore that the location of the occulting dust is as close to the star as physically
possible, being set by sublimation (e.g. \cite{2007prpl.conf..539M}). These systems tell
us about the nature of turbulence and accretion in gas-dominated disks, but so far reveal
little about how these disks transition to the debris phase and the subsequent evolution.

Here we focus on RZ~Psc, a star that shows UXor-like variability
(e.g. \cite{1985PZ.....22..181Z,1999A&AS..140..293G,2003ARep...47..580S}). As a young K0V
type star with no evidence for gas accretion and a strong infrared (IR) excess, this
system appears unique among UXors and may provide new information on the structure of
inner planetary systems during or following dispersal of the gas disk
\cite{2010A&A...524A...8G,2013A&A...553L...1D}. However, the IR excess indicates the over
5\% of the starlight is intercepted by the disk, which is a level more akin to gas-rich
protoplanetary and transition disks than debris disks. Specifically, we use a decade of
ground-based optical photometry of RZ~Psc (section \ref{s:data}) to draw conclusions on
dust location (section \ref{s:where}), and discuss the possible disk structure and
evolutionary state in section \ref{s:disk}. We conclude in section \ref{s:conc}.

\section{A clumpy dust ring near 0.5 au?}\label{s:rzpscintro}

There is significant evidence that the optical variations seen towards RZ~Psc are caused
by circumstellar dust: i) during dimming events the colour becomes redder
\cite{1985PZ.....22..181Z,1980PZ.....21..310K} in a way consistent with that expected for
dust \cite{1981Afz....17...87P,2004ARep...48..470P}, ii) the maximum depth is about
2.5mag and during these events the colour returns to near stellar values, suggesting that
the remaining emission is from light scattered off the circumstellar dust (i.e. the star
is fully occulted \cite{1981Afz....17...87P,1988SvAL...14...27G}, and iii) the
polarisation fraction increases during the transits, as expected if an increasing
fraction of the light is scattered off a disk of circumstellar dust
\cite{1988SvAL...14...27G,1991Afz....34..333K,2003ARep...47..580S}.

What separates RZ~Psc from other UXors (and dippers) is i) the spectral type is K0V
rather than Herbig Ae/Be for UXors and late K to M-type for dippers, ii) the occulting
dust lies well beyond the sublimation radius, and iii) the star is not associated with a
star-forming region so is inferred to be a few tens of millions of years old. The dust
distance has been inferred from the speed of ingress of dimming events, which was
previously estimated as about 0.6au (for circular orbits
\cite{2013A&A...553L...1D}). Corroborating evidence comes from the $\sim$500K temperature
of the dust seen in the mid-IR, which places it near 0.4-0.7au (depending on optical
depth) and therefore at a location consistent with the occulting dust
\cite{2013A&A...553L...1D}. The distance to RZ~Psc is unknown, but as an apparently
isolated star that shows Li absorption the age has been estimated as a few tens of Myr,
and therefore beyond the age at which a gas-rich disk would normally exist
\cite{2010A&A...524A...8G,2014A&A...563A.139P}. Further distinguishing features are that
the duration of the dimming events is consistently short compared to other UXors, a few
days rather than days to a few weeks, and that no near-IR (i.e. K-band) excess or
accretion signatures are seen \cite{2014A&A...563A.139P}, so interpretations related to
accretion of disk material onto the star
(e.g. \cite{1999AJ....118.1043H,1999A&A...349..619B,2016arXiv160503985B}) are unlikely.

Thus, the potentially compelling and unique aspect for RZ~Psc is that we are observing
dimming events from dust in a main-sequence planetary system that resides at about
0.5au. This dust is also seen in thermal emission, so deriving joint constraints on the
dust properties and structure may be possible. As argued by de~Wit et
al. \cite{2013A&A...553L...1D} a picture is emerging in which RZ~Psc is surrounded by a
massive young version of our own asteroid belt, in which planetesimals are continually
being destroyed. These collisions generate the large collective surface area of small
dust that emits strongly in the mid-IR, and the system geometry means that this dust also
sometimes passes in front of the star.

While this asteroid belt picture is intriguing, and makes RZ~Psc a system that could be
of great interest and worthy of detailed study, it is not the only possibility. Well over
1\% of the starlight is reprocessed by the circumstellar disk, which is more typical of
the primordial gas-rich disks seen around nearly all young stars. The discovery of
systems like HD~21997, that appear to be a few tens of Myr old and host gas-rich disks
\cite{2013ApJ...776...77K}, shows that stellar age is not a perfect indicator of disk
status. Thus, a considerable part of our analysis focuses on the question of the status
of the disk around RZ~Psc.

Given the proposed interpretation related to individual planetesimal disruptions, rather
than hydrodynamics, it is perhaps surprising that to date the dimming events are not seen
to be periodic \cite{1999AstL...25..243R,2013A&A...553L...1D}. The only cyclical
variation seen in light curves for RZ~Psc is a 12.4 year variation with an amplitude of
0.5 mag, which is attributed to either a magnetic cycle, or precession of an otherwise
unseen outer disk due to perturbations from an unseen companion
\cite{2013A&A...553L...1D}.

\section{Time series photometry}\label{s:data}

\subsection{Optical}\label{ss:opt}

\begin{figure*}
  \begin{center}
    \hspace{-0.5cm} \includegraphics[width=\textwidth]{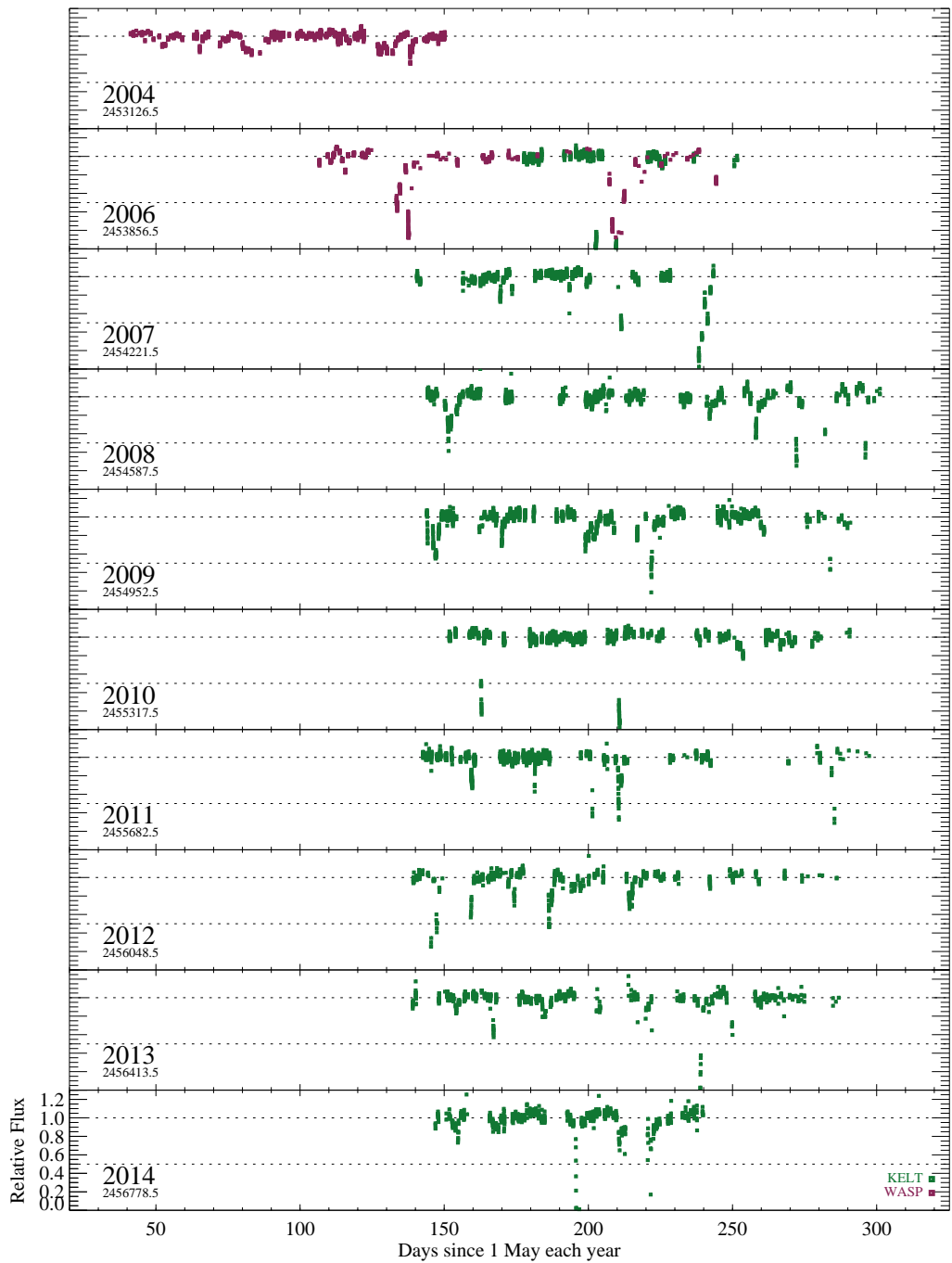}
    \caption{WASP and KELT-North data. Photometry is shown in dimensionless form,
      relative to a quiescent level of 1, and was converted from observed magnitudes as
      described in the text.}\label{fig:waspkelt}
  \end{center}
\end{figure*}

To study the temporal variability of RZ~Psc we use two seasons of public data from the
Wide-Angle Search for Planets (WASP \cite{2006PASP..118.1407P,2010A&A...520L..10B}), and
nine seasons of data from the Kilodegree Extremely Little Telescope North (KELT-North
\cite{2007PASP..119..923P}). We also collected, but ultimately did not use, photometric
observations of RZ~Psc from a wide variety of other sources
(\cite{1994AJ....108.1906H,1973IBVS..783....1K,1980PZ.....21..310K,1985PZ.....22..181Z,1991Afz....34..333K,1997AcA....47..467P,2014Ap.....57..491P},
the Catalina Sky Survey, the American Association of Variable Star Observers, the All-Sky
Automated Survey). Aside from the Harvard plate photometry published by G\"urtler et
al. \cite{1999A&AS..140..293G} we have not sought unpublished photometry so the light
curve remains incomplete.\footnote{All normalised photometry is available at
  \href{https://github.com/drgmk/rzpsc}{https://github.com/drgmk/rzpsc}}

Here we focus on the WASP and KELT-North data, as it has not been previously analysed and
has considerably higher cadence (many measurements per night) and temporal coverage
(nightly, weather permitting) than other data sets. The WASP data from 2004 and 2006 are
public and were obtained from an online
archive\footnote{\href{http://wasp.cerit-sc.cz}{http://wasp.cerit-sc.cz}}. These data
were processed in a manner similar to that described by \cite{2014MNRAS.441.2845V}, where
common-mode variations were removed using 50 quiet nearby stars. The WASP bandpass is
broad, with roughly uniform transmission from 400-700nm \cite{2006PASP..118.1407P}. The
KELT-North data, 2006-2014, were used in raw form, the only specific treatment being a
4\% relative correction being made to ensure observations taken in the ``east'' and
``west'' telescope orientations have the same calibration. The bandpass is redder than
for WASP, with most transmission between 500nm and 800nm \cite{2007PASP..119..923P}.  For
a full description of the KELT-North data reduction, see \cite{2012ApJ...761..123S}.

We normalised each year's data from each instrument separately by converting magnitudes
to flux density and dividing out the sigma-clipped median so the light curve has an
out-of-occultation baseline of 1. In doing so we are assuming that variations due to the
slightly different filter bandpasses are unimportant. Each row in Figure
\ref{fig:waspkelt} shows a season's data, starting on May 1 each year (JD also
indicated). Most year's data therefore extend into the next year, so the ``2006 data''
refers to data from the 2006/2007 observing season.

\subsubsection{Qualitative light curve overview}\label{sss:quallc}

\begin{figure*}
  \begin{center}
    \hspace{-0.5cm} \includegraphics[width=\textwidth]{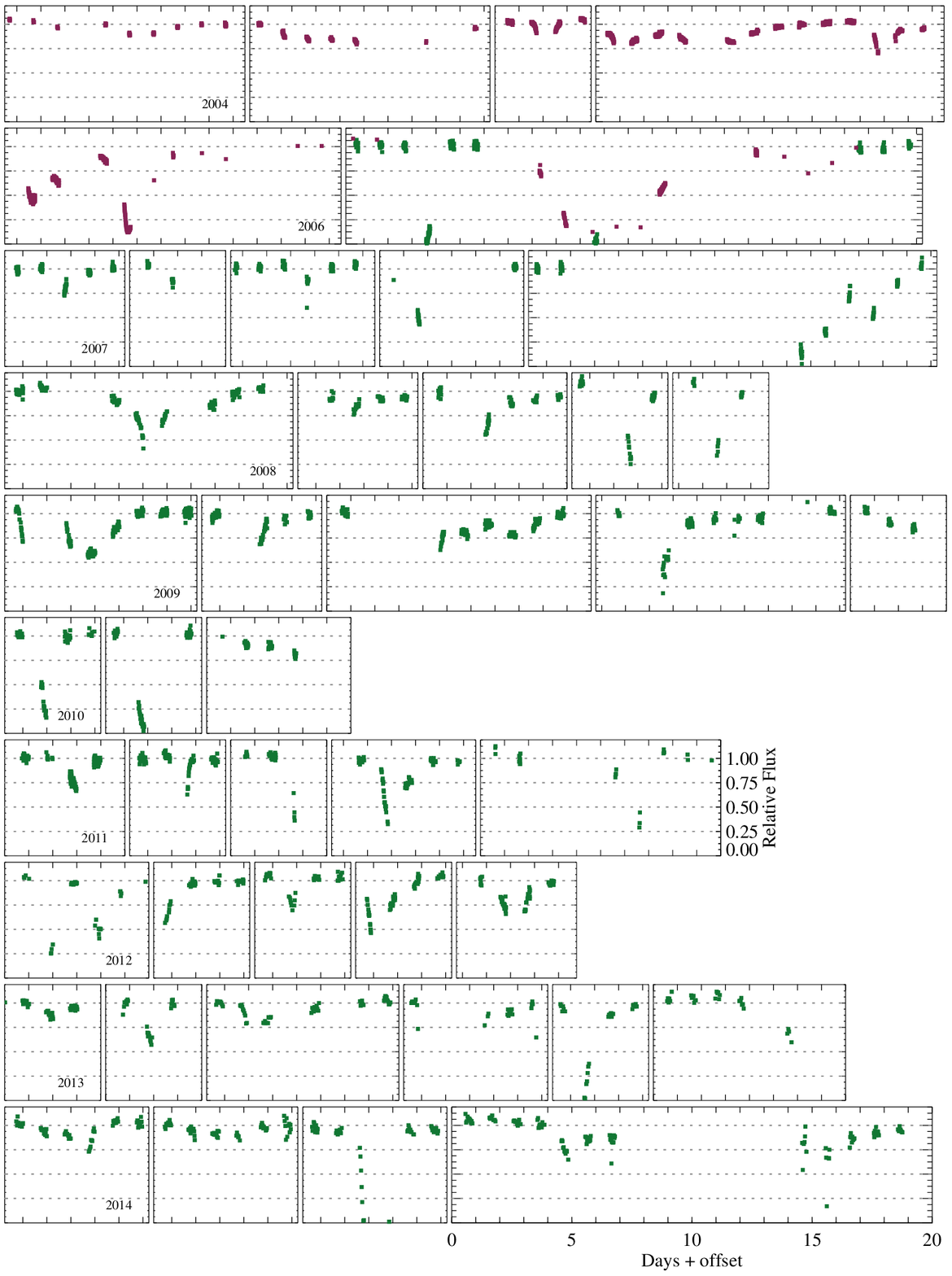}
    \caption{WASP and KELT-North data, focussing on dimming events. The vertical and
      horizontal scales in each sub-panel are the same.}\label{fig:waspkeltzoom}
  \end{center}
\end{figure*}

It is clear from Figure \ref{fig:waspkelt} that RZ~Psc undergoes the very deep dimming
events that are typical of UXors. These are seen a few times each observing season and
vary in complexity, with a few extended events (e.g. 2006) and a greater number of
``neater'' single events (e.g. 2010). In some years there is also significant variability
at shallower depths. Of particular note is the pair of deep events in 2006; these appear
to be about 70 days apart, and given the suggestion that the putative asteroid belt
analogue resides near 0.5au a natural inference is that these two events are related. If
true, this repetition corresponds to a semi-major axis of about 0.3au, which given
uncertainties in the true disk spectrum could be consistent with the location of the
asteroid belt. In 2004 there are about 100 days of near-consecutive nights of data and no
deep events, so either the true period is longer than 100 days, or dust clumps can be
created (and perhaps destroyed) on timescales of a year or so.

In Figure \ref{fig:waspkeltzoom} we have selected most of the events from each year and
shown them at a greater temporal resolution. The scale in each panel is the same, so
wider boxes simply cover longer events. Most events appear to last at least a few days,
suggesting that only having nightly coverage does not seriously hinder our ability to
detect most events. However, the events are sufficiently short and irregular that the
true shape of events remains uncertain. While it is likely that interpolation of the
photometry for the fourth event in 2011 (that is, the fourth box from the left in the row
corresponding to 2011 in Figure \ref{fig:waspkeltzoom}) would resemble the true light
curve, this assumption seems very unlikely to yield the true evolution of more complex
events like those in 2006.

Nevertheless, Figure \ref{fig:waspkeltzoom} shows an unprecedented view of dimming events
seen towards RZ~Psc, and that key information on the ingress and egress of dimming events
is present. The dimming rate is such that it can be resolved temporally, and hence the
velocity and radial location of the dust clumps estimated. While such estimates have been
made in the past based on one or two individual events \cite{1985PZ.....22..181Z}, these
data make them possible for an ensemble of tens of events.

A fairly basic question is whether the light curve could result from objects that all
have the same properties, or whether a range is required. Given the existence of both
long shallow events and short deep events, at a minimum the clumps must vary in size
and/or velocity across the face of the star, but probably also have different optical
depths. The star can be completely occulted, so the clumps can be optically thick and
star-sized, but it has already been shown that the events are not grey in colour, so the
clumps must have a density gradient rather than sharp edges. Where sufficient data exist,
it is clear that not all events have the same relative shape, so the shape of the clumps
varies. Thus, the broad picture is of roughly star-sized clumps, whose shape and orbital
elements vary. The fact that the dimming events can be shown in a series of panels with
the same scale in Figure \ref{fig:waspkeltzoom} suggests that the range over which these
properties vary is of order factors of a few, not many orders of magnitude.

\subsection{Infrared}\label{ss:irvar}

While the optical photometry reveals information about how RZ~Psc is itself dimmed, IR
photometry beyond a few microns is dominated by emission from the circumstellar
disk. Thus, IR variation reveals information about how the emitting surface area,
temperature, and perhaps composition, of the dust change with time. Such variation is
indeed apparent, both from comparison of an AKARI 18$\mu$m non-detection at a lower level
than the WISE 22$\mu$m detection, and from several individual WISE measurements taken at
6-month intervals.

\begin{figure*}
  \begin{center}
    \hspace{-0.5cm} \includegraphics[width=0.7\textwidth]{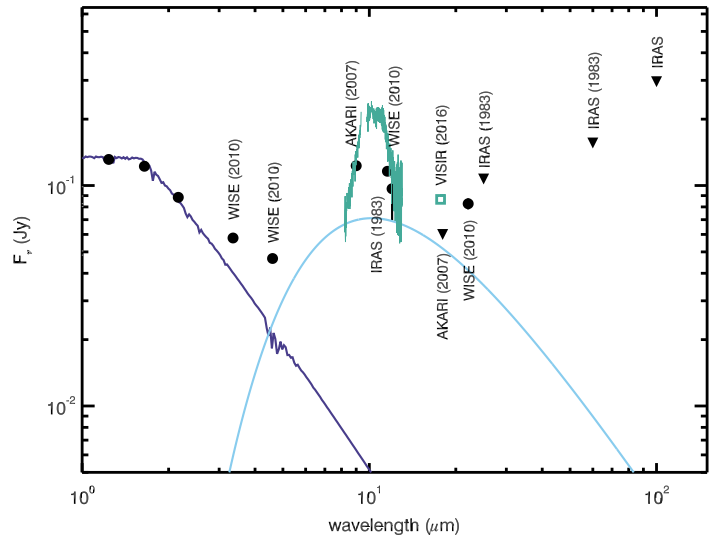}
    \caption{Flux density distribution of RZ~Psc, including 2MASS, WISE, AKARI, VISIR,
      and IRAS data and their (approximate) year of observation. The dark blue line shows
      a stellar photosphere model at the approximate stellar temperature of 5350K, and
      the light blue line a 500K blackbody. The latter is not a fit, but an approximate
      continuum level that illustrates that the WISE 3.4 and 22$\mu$m photometry cannot
      both be accounted for with a single blackbody, if the silicate feature seen with
      VISIR was present in 2010.}\label{fig:sed}
  \end{center}
\end{figure*}

Motivated by this variation, we obtained VLT/VISIR observations of RZ~Psc; an N-band
(10$\mu$m) spectrum on 2016 August 16 and Q-band (18$\mu$m) photometry on 2016 July 27
(programme 097.C-0217). These data, and the related calibration observations, were
reduced using the standard ESO \texttt{esorex} pipeline. The wall-clock integration time
for the spectrum was 50 minutes at an airmass of 1.66, and was calibrated using an
observation of HD~189831 taken immediately afterwards at an airmass of 1.63. The spectra
in individual chop/nod cycles are consistent so the shape of the spectrum is
reliable. The absolute calibration is uncertain at $\sim$10\% levels
\cite{2007A&A...476..279G}, which is sufficiently precise for our purposes here. The
spectrum was trimmed to mask highly uncertain regions shortward of 8$\mu$m, and longward
of 13 $\mu$m, and near the telluric absorption at 9.5$\mu$m. The Q-band photometry took
45 minutes at an airmass of 1.65-1.7, and was calibrated against an observation of
HD~2436 taken immediately afterwards at an airmass of 1.5. Photometry of RZ~Psc and
HD~2436 was done using a 0.9'' radius aperture and a sky annulus from 1-2''. In addition
to the conversion from adu/s to Jansky using HD~2436, we applied an additional upward
correction to account for the slightly lower airmass for the calibrator
($\exp(0.3 [1.675-1.5] \approx 1.05$, where an extinction of 0.3 per unit airmass was
used, derived from archival calibration data using a method similar to that of Verhoeff
et al. \cite{2012A&A...538A.101V}). Uncertainties were estimated as the standard
deviation of the flux density in 50 apertures around RZ~Psc. The final calibrated flux is
$86 \pm 10$ mJy.

The VISIR spectrum and photometry are shown in Figure \ref{fig:sed}, which also shows the
other available near- to far-IR photometry. The absolute level of the spectrum agrees
well with the IRAS, WISE (from the ALLWISE catalogue), and AKARI observations; that four
observations spanning over 30 years are consistent suggests that while this part of the
spectrum may vary, the shape and level shown is typical. The same cannot be said near
18$\mu$m, where the AKARI and VISIR (and perhaps WISE depending on the disk spectrum)
flux densities are inconsistent. The flux near 10$\mu$m being relatively constant, and
the 18$\mu$m flux changing could be indicative of shadowing of an outer disk by an inner
disk; that is, evidence that the disk around RZ~Psc has significant radial extent. This
type of behaviour is seen as ``seesaw'' variability in IR spectra towards some transition
disks (e.g. \cite{2011ApJ...728...49E,2012ApJ...748...71F}).

The spectrum clearly shows solid-state emission (and the continuum level is not actually
clear), which indicates i) that the dust is of order microns in size, and ii) that the
dust includes silicates. The smooth rise and fall suggests that the silicates are largely
amorphous; crystalline silicates have sharper features, notably a depression between two
peaks at 10 and 11 $\mu$m, rather than the flat top seen here. Other systems thought to
host bright asteroid belt analogues (rather than gas-rich disks), such as HD~69830,
BD+20~307, and HD~113766A, tend to show crystalline features
(e.g. \cite{2005ApJ...626.1061B,2005Natur.436..363S,2012A&A...542A..90O}), which may
argue against such a scenario for RZ~Psc. However, such comparisons are largely
speculative as there is also a high degree of variation among silicate features, for both
gas-rich and gas-poor disks.

\begin{figure*}
  \begin{center}
    \hspace{-0.5cm} \includegraphics[width=\textwidth]{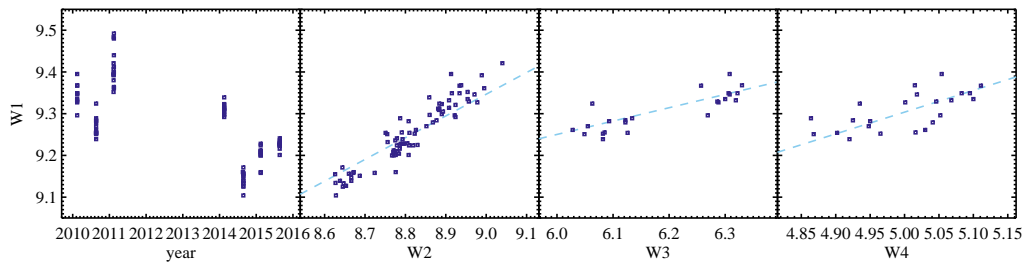}
    \caption{WISE epoch photometry at 3.4, 4.6, 12, and 22$\mu$m (W1, W2, W3, and W4, in
      magnitudes). The left panel shows the time variation in W1 over 5.5
      years. Subsequent panels show how W1 correlates with W2, W3, and W4, which do not
      have observations at all W1 epochs. The dashed lines show the slope expected for
      constant disk flux variation with wavelength (the variation is smaller in W1/2
      because the total flux is not dominated by the disk).}\label{fig:wisevar}
  \end{center}
\end{figure*}

To further explore this variability, we use the WISE ``epoch'' photometry, for which the
telescope scanning strategy results in clusters of measurements that are spaced 6 months
apart. These data appear at approximately days 70 and 250 in the relevant years in Figure
\ref{fig:waspkelt}, but do not coincide with any optical dimming events. Photometry is
not available in all four channels since launch in early 2010, due to exhaustion of the
coolant after 7.7 months and a 2.5-year hiatus from mid 2011-2014 (see
\cite{2010AJ....140.1868W,2014ApJ...792...30M}). These data are shown in Figure
\ref{fig:wisevar}, where the left panel shows the 3.4$\mu$m magnitude as a function of
time, and that there is significant variation on 6-month timescales, and an even greater
variation overall. Inspection of the individual clusters, which are on hour to day
timescales, shows no significant variation with time. The dashed lines have the slope
expected for disk brightness variation that is independent of wavelength; the slopes are
not exactly 1 because the total flux is not dominated by the disk near 3-5$\mu$m, and
hence the slopes are slightly flatter. Comparing the observed and expected correlations,
we conclude that the data do not show significant evidence for changes in the spectral
shape (i.e. changing temperature or composition). However, the ratios including 12 and
22$\mu$m observations are most sensitive to these changes, but only include the first two
sets of observations where the brightness changes were relatively small.

Nevertheless, the amplitude of the change in 3.4$\mu$m brightness over 5 years is about
30\%. Considering that the disk flux density is only 45\% of the total flux at this
wavelength, the disk brightness increased from 2010 to 2015 by about a factor of two. A
similar variation can be inferred by comparing the 18$\mu$m upper limit from AKARI in
2007 and our VISIR measurement in 2016. Given these increases is it surprising that the
N-band spectrum does not appear much higher than the IRAS, AKARI, and WISE photometry. A
possible explanation would be that the increased emission originates in larger grains,
which would result in greater continuum flux but similar levels in spectral
features. However, without wider spectral and more frequent temporal coverage quantifying
such effects is difficult. This level of IR variation is seen towards both protoplanetary
(e.g. \cite{2011ApJ...728...49E,2014ApJ...791...42Z}) and debris disks
(e.g. \cite{2012ApJ...751L..17M,2012Natur.487...74M,2014Sci...345.1032M}), so these data
provide little means to distinguish between scenarios.

\section{Where are the occulting bodies?}\label{s:where}

The main part of our analysis concerns attempts to extract information from the optical
light curve, taking advantage of the great number of dimming events seen over ten
seasons. In this section we focus on the radial location of the bodies (``clumps'') that
pass in front of RZ~Psc, first searching for periodicity associated with repeat events,
and then using the light curve gradients to constrain the projected velocities. This
analysis primarily focusses on what can be gleaned from the light curves, and the
implications of these results for different clump origins are then explored in section
\ref{s:disk}.

\subsection{Search for periodic dimming events}\label{ss:per}

We begin by estimating the lifetime of an occulting clump as a check on the plausibility
that dimming events should repeat. The angular rate at which clumps are sheared
  out is R$d \Omega / dR = - 3 \Omega / 2$. Accounting for shear in both forward
  (interior) and backward (exterior) directions the shear velocity across a clump of
  radius $R_{\rm cl}$ is then
\begin{equation}\label{eq:shear}
  v_{\rm sh} = 3 R_{\rm cl} \Omega \, ,
\end{equation}
so the clump expansion rate due to shear in units of clump radii is only three times the
orbital frequency. That is, after one orbit a clump will be stretched by a factor of
$6\pi$, and the radial and vertical optical depth will be roughly $6\pi$ lower (though it
might still be optically thick). Thus, clumps that are not bound by their own
self-gravity are expected to have a short lifetime at optical depths that are large
enough to cause detectable dimming events, but could cause repeated dimming events if
they are initially optically thick.

The temporal coverage of the observations in an individual season is 100-150 days. Thus,
if the occulting material resides in an asteroid belt closer than $\sim$0.5au,
periodicity in the dimming events may be visible in a single season's data. Longer
orbital periods may be visible across seasons, though the six-month gap between seasons
makes unambiguously linking events harder. Non-detection of periodicity would imply that
the data are not sufficient, or that strict periodicity does not exist. An intermediate
possibility is that occultations happen with a range of periodicities, perhaps reflecting
their origin in a radially broad region, and that discerning this scenario from randomly
occurring occultations is not possible given the data.

In an attempt to find the expected periodicity we tried several approaches. These are
similar in that they aim to quantify whether some feature in the light curve is repeated
again at a later time, but differ in how well they reveal evidence for a periodic
signal. We found no evidence for events that are related from one year to another, so
focus on statistics derived from individual seasons' data (though these are sometimes
combined).

\subsubsection{Autocorrelation}\label{sss:auto}

\begin{figure}
\begin{center}
\hspace{-0.5cm} \includegraphics[width=0.7\textwidth]{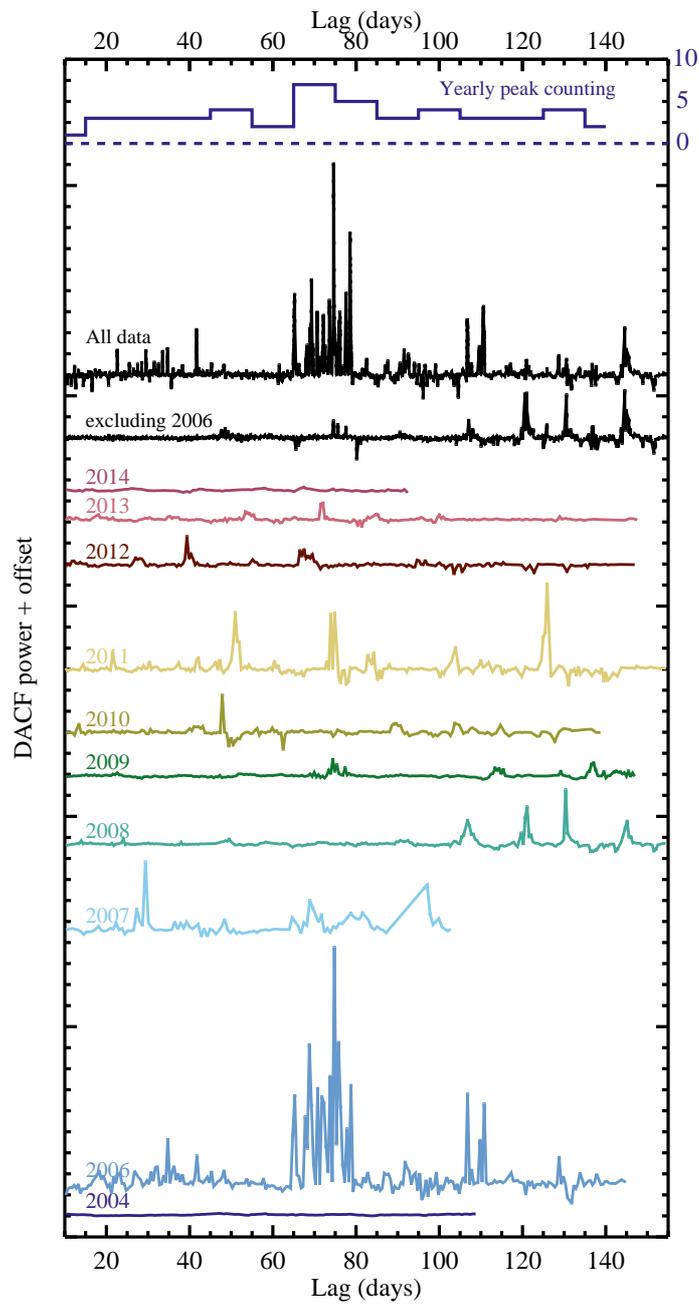}
\caption{Discrete autocorrelation function for yearly WASP and KELT-North data, computed
  for lags between 10-155 days). The second and third lines from the top show all data,
  and 2006-excluded data. The topmost line shows the number of years that show a peak
  more than 3$\sigma$ above the clipped DACF mean within each 5 day bin. The peak at 70
  days is 7 years, the dashed line is zero, and the y-axis scale is shown to the
  right.}\label{fig:auto}
\end{center}
\end{figure}

We first used autocorrelation to search for periodicity, rather than methods related to
Fourier transforms (e.g. periodograms). The motivation being that an individual transit
event may be followed by another some number of days later, and perhaps repeat a few
times, but other similarly (but not exactly) separated events may happen years later or
earlier with a phase that is totally different. We therefore used the discrete
autocorrelation function (DACF) proposed by Edelson \& Krolik \cite{1988ApJ...333..646E},
though do not include uncertainties on individual measurements. For a time series with
measurements $a_i$ at times $t_i$ the DACF first computes the mean $\bar{a}$ from the
light curve, here we used the sigma-clipped mean to remove the dimming events. Then for
each pair of points $a_i$, $a_j$ (with $i\ne j$) computes
$U_{ij}=((a_i-\bar{a})(a_j-\bar{a}) / \sigma_a$, with each $U_{ij}$ associated with a
time lag $\Delta t_{ij}=t_j - t_i$. A series of time lags centered at times $t_{\rm lag}$
with width $\Delta t_{\rm lag}$ are then used as bins, and the average in each bin is the
DACF. The DACF is not computed for lag bins with no data. The units of the DACF are
standard deviations of the light curve $\sigma_a$ (again calculated using sigma
clipping).

The results are shown in Figure \ref{fig:auto} for time lags (i.e. trial periods) of 10
to 155 days in half-day bins. Comparison of these with the light curves shows that the
DACF recovers most, but not all, events. Conversely, not all DACF peaks are necessarily
associated with real repeat events, as there may of course be multiple distinct clumps
orbiting the star at any given time. Not all pairs of events show a strong DACF signal,
as they can comprise only a few measurements and the mean for that $t_{\rm lag}$
dominated instead by a much larger number of measurements elsewhere in the light curve
closer to the quiescent level (i.e. near $\bar{a}$). Our attempts to avoid this issue by
using autocorrelation on interpolated data yielded mixed results; heavy filtering, such
as setting all data above a given level to 1, was needed for results similar to the DACF
shown in Figure \ref{fig:auto}.

While several strong peaks appear in the DACF of all data, most of these arise from 2006,
as can be seen in the DACF when these data are excluded. Some peaks remain near 70 days,
as well as at 120 and 145 days, and the latter two could be aliases of periods near 60-70
days, arising simply because an event was missed. That is, the irregular sampling means
that absence of evidence of power at some period in the DACF is not evidence of absence.

The pair of events separated by 70 days in 2006 provides the strongest signal, and most
other years also show events near this period (2007, 2009, 2011, 2012, 2013, 2014). To
illustrate these numbers the topmost line in Figure \ref{fig:auto} quantifies the number
of years that show a peak, in 10 day bins. The peak of 7 years is at 65-75 days, which is
suggestive but not conclusive because a K-S test shows that this distribution is
consistent with being uniform in period.

\subsubsection{Iterative event finding}\label{sss:iter}

In an attempt to avoid some of the difficulties arising from the DACF, we tried a similar
approach that first identifies individual occultation events and then computes the time
delays between them. The main aim was to identify and use all events in a way that avoids
biases related to the sampling of the data and the different relative depths of
potentially repeated events. By repeating this prescription for synthetic data we are
able to test different scenarios for how the occultations do or do not repeat.

For this approach, an event is initially identified as the lowest point that is 6$\sigma$
below the mean, where the mean and standard deviation are again estimated by
sigma-clipping. This lowest point is noted, as are all immediately adjacent points that
are also below the threshold noted. The points so included constitute a single dimming
event. The points belonging to this event are removed from the light curve and the
process repeated until no significant events remain. The time of the event is the time of
the lowest point, and the duration the time between the two end points that are
consistent with the quiescent level. Thus, if an event is in a region of sparse sampling
the duration can appear to be longer than it probably is, though we discard any events
that occur at the beginning or end of an observing season to avoid unreasonably long
events.

For a given set of event times and durations, the range of possible times between events
is then calculated using the maximum allowed by the duration. This calculation is done
for all combinations, yielding $N(N-1)/2$ inter-event time ranges. These ranges are then
``stacked'' into a histogram (i.e. counting +1 for time differences within a given bin)
that shows a measure of the power present at a given time difference.



\begin{figure}
  \begin{center}
    \hspace{-0.5cm} \includegraphics[width=0.7\textwidth]{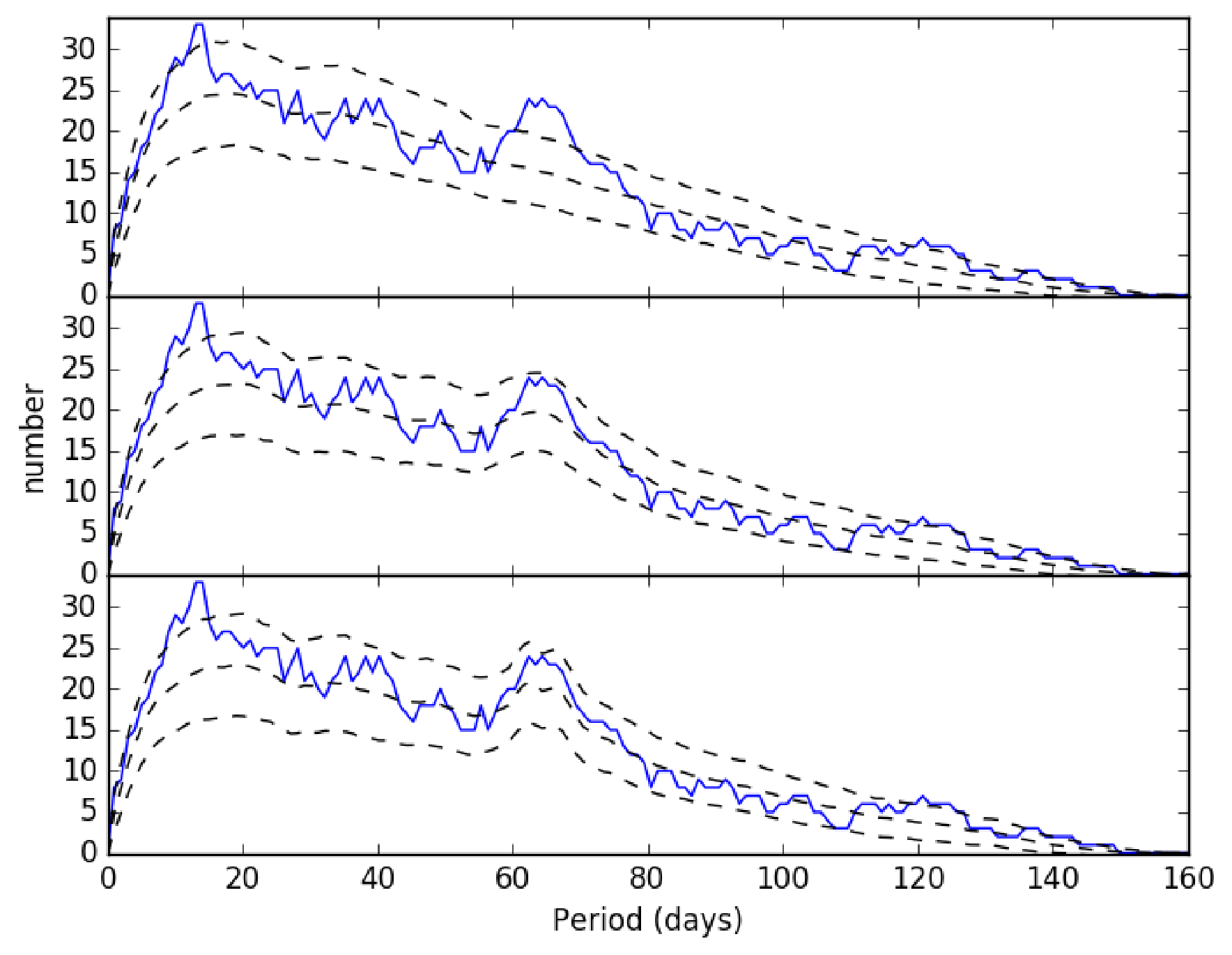}
    \caption{Power at a given period from the iterative event finding. The solid line
      shows the power from the data, and is the same in each panel. The dashed lines show
      the mean and $\pm$1$\sigma$ power from simulated dimming events, which from top to
      bottom are: random, periodic between 64 and 75 days with 3 repeats, and periodic at
      64 days with 10 repeats.}\label{fig:hist}
  \end{center}
\end{figure}

The solid lines in Figure \ref{fig:hist} show this power in histogram form (the same in
each panel), generated from 60 events that were identified in the light curve. The y-axis
should be interpreted as the number of events that are consistent with that period on the
x-axis. As before, there is evidence for a peak, now slightly shifted to near 65 days. In
contrast to the DACF analysis, this power does not all arise from a single year. Most is
contributed by 2004 and 2009, but exclusion of events from these years results in a
similar (but noisier) histogram.

To test what could have been detected, and quantify the variation in power expected,
we created synthetic light curves with the same temporal sampling as the WASP and
KELT-North data. To do this we set the flux to 1, and randomly injected a number of
dimming events with a flux of 0.1. These events are therefore relatively easy for the
algorithm to detect, but suffer the same sampling issues. We tried individual randomly
occurring events, and periodic events that repeat a fixed number of times. All events
have durations randomly distributed between 1 and 4 days, similar to the observed
events. For random events the remaining parameter in this model is simply the number of
events -- this is the total number over the 11 year period covered by the WASP and
KELT-North data, including when measurements were not being taken. For repeating events
the range of periods and the number of repeats are additional parameters. To estimate the
level of variation in the power spectrum we repeated the process of injecting synthetic
events 500 times, and in each bin estimate the mean and standard deviation of the power.

The results are shown in Figure \ref{fig:hist}, where the dashed lines show the mean and
$\pm$1$\sigma$ deviations from the simulations (and solid lines show the data). Each
panel shows a different scenario, from top to bottom these are: 295 random events, 100
events with periods between 64 and 75 days and 3 repeats, and 100 events with a period of
64 days and 3 repeats. The purely random events are marginally disfavoured; while the
data lie outside the dashed lines, these are only 1$\sigma$. Nevertheless, the middle and
bottom panels show that the possible peak near 65 days could be caused by events with a
range of periods, but that only a single period of 64 days is actually needed and yields
a slightly stronger peak. The remaining excess periodicity near 10 days is not accounted
for by any of the models, but could be an indication that on this timescale events are
related (e.g. a clump that has separated into several). That the real signal nearly lies
within the 1$\sigma$ variation expected for randomly occurring events shows that the
evidence for periodicity is weak, but that as found by the autocorrelation analysis, this
weak evidence points towards 64 days as a possible period.

A side-effect of the simulations is an estimate of about 300 dimming events in total over
the period between 11 June 2004 and 21 February 2015 (3851 days). On average an event
therefore occurs about once every 13 days and there are 30 dimming events each year (and
roughly 15 per observing season). If events repeat three times, then a \emph{new} dimming
event would appear every 39 days on average, but we would also expect to see two other
unrelated events during this time.

The total is five times larger than the 60 events detected by the iterative search. Most
of these events were therefore either missed by the WASP and KELT-North observations, or
not counted because they were mis-identified as single events. From Figure
\ref{fig:waspkelt} a rough estimate is that 3 in 5 events would have been missed due to
incomplete temporal coverage, meaning that about 1 in 5 simulated events were
sufficiently close to other events that they were not separately identified. Not all
events are deep, but as could be surmised from Figure \ref{fig:waspkelt}, near continuous
observation of RZ~Psc would yield a rich light curve. If continuous coverage allowed
closely occurring events to be separated, the number identified in a given time period
would approximately double.

\subsubsection{Summary of period search}\label{sss:persum}

We found a weak periodic signal near 60-70 days, but neither of the methods described
above show compelling evidence that the dimming events seen towards RZ~Psc are periodic
and not random. While we presented results for single seasons' data, we saw no evidence
for periodicity on longer periods. Aside from the 12 year variation, no periodicity has
been seen in the past \cite{2013A&A...553L...1D}. These searches used periodograms,
which are sensitive to variations with fixed phase and poorly motivated, so we explored
autocorrelation and a similar method. That we found a possible signal can be attributed
to both a different method, and the significantly better temporal coverage of the WASP
and KELT-North data.

A lack of strong evidence for periodicity is perhaps surprising, since material that
occults the star once and is on an unperturbed orbit must pass in front of it again. Not
all material need return at the same time however, and the prediction of the shearing
estimate made at the outset, that the visible lifetime of clumps when they are optically
thin is similar to the orbital period, appears to be borne out. Of course, shearing is
not the only possible explanation, as pressure effects in a hydrodynamic turbulence
scenario might also disperse a clump (as the sound crossing time for a star-sized clump
near 1au is of order or shorter than an orbit). The latter scenario relies on a
significant gas reservoir, so the primary test to distinguish between different clump
scenarios lies with the evolutionary status of the disk, which we explore in section
\ref{s:disk}\ref{ss:evol}.

\subsection{Gradient analysis}\label{ss:grad}

Given the possibility of a 60-70 day periodicity, the location of the occulting bodies
could be relatively close to the star, with semi-major axes of about 0.3au. This distance
is comparable to the 0.4au estimated for optically thin dust at 500K
\cite{2013A&A...553L...1D}. To further investigate the location we turn to a different
aspect of the light curves that provides information on the velocity of the occulting
bodies; the gradients. To convert gradients measured in the light curves to velocity and
orbital distance, we first outline a simple model, and then use this model to interpret
the data.

\subsubsection{``Curtain'' model}\label{sss:curtain}

This section considers a simple one-dimensional model (along $x$) of a cloud that dims a
star. The main assumption is that the cloud is larger than the star, so for a cloud that
passes in front of the star from left to right, the vertical ($y$) size of the cloud can
be ignored. The large cloud extent is suggested primarily by the large depths of the
dimming events, but also because no flat-bottomed (i.e. planet transit-like) events are
seen. It seems likely that not all clumps are this large, and that a variety of sizes
(and impact parameters) exist, but for our purposes this simplification is
sufficient. Thus, the cloud is modelled as a semi-opaque screen or ``curtain'' that dims
the star, as in previous analyses of related phenomena (e.g. KH-15D, J1407
\cite{2006ApJ...644..510W,2014MNRAS.441.2845V,2015ApJ...800..126K}).

The star is dimmed by the passage of a cloud located at $x_{\rm cl}$ from the star
center. The 1-D geometric optical depth structure of the cloud is given by some function
centred at $x_{\rm cl}$ (e.g. a top hat or Gaussian) so is $\tau(x-x_{\rm cl})$. The star
has a surface brightness $I(\sqrt{x^2+y^2})$, which could allow for limb-darkening. The
observed flux from the star is then
\begin{equation}\label{eq:f}
  F(x_{\rm cl}) = \int_{-R_\star}^{R_\star} (1-\tau(x-x_{\rm cl})) dx
 \int_{-\sqrt{R_\star^2-x^2}}^{\sqrt{R_\star^2-x^2}} I(\sqrt{x^2+y^2}) dy \, ,
\end{equation}
which first integrates vertically over the star at some $x$ (i.e. independently of
$\tau$), and then along $x$, which includes the effect of the cloud. The light curve is
therefore the convolution of the 1-D stellar brightness profile with the clump's optical
``thin-ness'' profile (i.e. $1-\tau$). The flux profile (light curve) is a function of
time, but the star and clump profiles are functions of $x$, and the conversion that links
these is the cloud velocity.


The simplest case is a star of uniform surface brightness that is occulted by an
optically thick screen that covers the star from $x=-1$ to $u$ (i.e. the units of length
are now $R_\star$). Then $I=1$ and $\tau(x-x_{\rm cl})$ is a step function at $u$ and the
fraction of the total stellar flux ($F_\star = \pi$) seen is \cite{2006ApJ...644..510W}
\begin{equation}
  f = F/F_\star = (cos^{-1}[u] - u\sqrt{1-u^2})/\pi \, ,
\end{equation}
where $f$ has the same units as our normalised light curve. If the curtain is not
completely optically thick then the fraction is instead
\begin{equation}
 f = F/F_\star + (1-F/F_\star)(1-\tau)
\end{equation}

The gradient of the normalised light curve as the curtain is pulled across is $df/du$,
and therefore the sky-projected (i.e. minimum) velocity of a clump is
\begin{equation}
  \frac{du}{dt} = -\frac{\pi}{2 \tau \sqrt{1-u^2}}\frac{df}{dt}
\end{equation}

A cloud that is not completely optically thick has a shallower flux gradient because it
reaches a shallower depth for the same velocity, and the factor $1/\tau$ accounts for
this effect. Stated another way, for this curtain model the optical depth and cloud
velocity are degenerate in producing some flux gradient. However, this degeneracy can be
partially broken because some information on $\tau$ exists; $\tau$ must be greater than
the depth of the dimming event (i.e. $1- f_{\rm min}$, the minimum normalised
flux). Objects somewhat smaller than the star are also accounted for; an optically thick
clump that covers half the star produces approximately the same light curve as a
$\tau=0.5$ clump that covers the whole star.

These expressions can be further simplified by assuming that the maximum gradient occurs
as the cloud ``edge'' passes the center of the stellar disk (i.e. $u=0$). By adopting a
radius for the star this velocity can be converted to physical units, and assuming a
stellar mass and that the clump is on a circular orbit converted to a semi-major axis. If
$df/dt$ is in units of fractional stellar flux per day (i.e. the light curve is
normalised and has time units of days) and $du/dt$ in stellar radii per day (i.e. units
of $u$ are $R_\star$), then the numbers for these quantities can be used in the following
equations:
\begin{equation}\label{eq:vcl}
  v \approx  8050 \frac{R_\star}{R_\odot} \frac{du}{dt} ~ {\rm m~s}^{-1}
\end{equation}
and
\begin{equation}\label{eq:acl}
  a_{\rm circ} 
  = 14 \frac{M_\star}{M_\odot} \left( \frac{R_\odot}{R_\star} \frac{dt}{du} \right)^2 ~
  {\rm au}
  = 8.7 \frac{M_\star}{M_\odot} \left( \tau \frac{R_\odot}{R_\star} \frac{dt}{df} \right)^2 ~
  {\rm au}
\end{equation}
While the assumption of $u=0$ yields a simple conversion between the light curve gradient
and the velocity and semi-major axis, it is of course possible to measure gradients that
are not at $u=0$. For example, the gradient when a dimming event reaches minimum is zero,
which implies zero velocity and an infinite semi-major axis (e.g. the first minimum in
2006 in Figure \ref{fig:waspkeltzoom}). In addition, orbits may not be circular and thus
the actual velocity is greater than the sky-projected velocity during a dimming
event. Thus, the gradients and velocities must be taken as lower limits, and the
semi-major axes as upper limits.

\subsubsection{Curtain model application}\label{sss:gradapp}

\begin{figure}
  \begin{center}
    \hspace{-0.5cm} \includegraphics[width=0.7\textwidth]{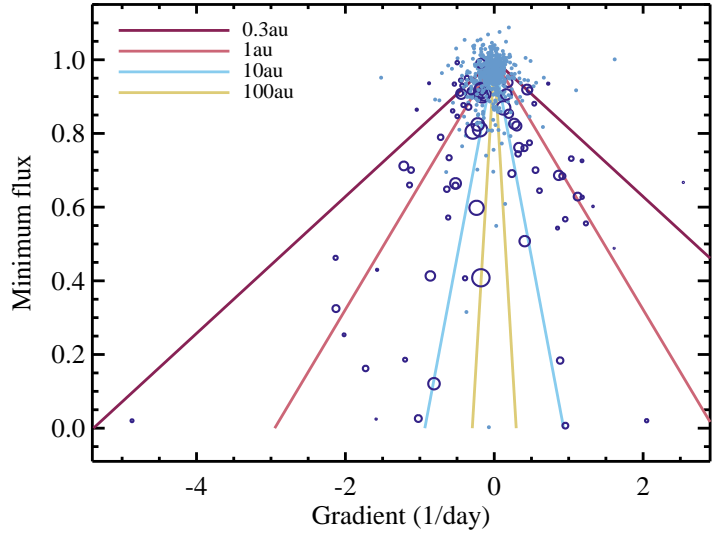}
    \caption{Gradient and minimum flux measured from individual nights'
      observations. Open circles have gradients significantly different from zero, and a
      symbol size proportional to the inverse of the gradient uncertainty. Dots are
      consistent with zero slope. The lines show the gradients implied by the velocities
      for circular orbits at 0.3, 1, 10, and 100au.}\label{fig:grad}
  \end{center}
\end{figure}

Using the simple formalism described above, we can use gradients derived directly from
the light curves to estimate the radial location of the occulting clumps. The gradients
are estimated by least squares fitting straight lines to each night's observations, for
which only nights with six or more measurements are used. This procedure is possible
because in nearly all cases an individual night's observations only cover ingress or
egress, not both. These gradients are plotted against the minimum nightly flux
$f_{\rm min}$ in Figure \ref{fig:grad}. We plot gradients whose uncertainty is less than
4$\times$ their value as open circles, with symbol sizes proportional to the inverse of
this uncertainty. All other gradients are plotted as small dots, as a check that the
gradients for unocculted fluxes are near zero. The horizontal scatter of these gradients
near $f_{\rm min}=1$ provides a further estimate of the uncertainty of individual
gradients. The solid lines show the gradients expected for circular orbits at a range of
semi-major axes.

An unusual feature in Figure \ref{fig:grad} is that the gradients may be biased towards
negative values at low $f_{\rm min}$. For the 15 negative, and 4 positive gradients below
$f_{\rm min}=0.5$, and approximately equal numbers of positive and negative gradients
above, Fisher's exact test yields a p-value of 0.02. Thus, there is evidence that the
distributions of gradients above and below minimum fluxes of 0.5 are different, with
negative gradients more commonly seen for deep dimming events. In the range
$0.5<f_{\rm min}<0.8$ there are 15 and 25 negative and positive gradients, a reversal of
the trend, but this difference only has a p-value of 0.1. Thus, there are about equal
numbers of gradients measured with minimum fluxes below 0.8, but their distributions are
different.

The bias to negative gradients below $f_{\rm min}=0.5$ suggests that ingress tends to be
slower than egress; the egress is too quick to be caught. However, the equal numbers
between $0.5 < f_{\rm min} < 0.8$ suggests that the rapid egress does not return the
light curve to quiescence, just to a level above $f_{\rm min}=0.5$. Thus, the statistics
suggest that a typical deep dimming event has an ingress at a rate of $-1$ to $-2$
day$^{-1}$, after which the flux rapidly rises to $f_{\rm min} \approx 0.5$, and then the
remaining egress is at a rate similar to ingress.

Qualitatively, this inference is consistent with a scenario of a disrupted asteroid,
whose structure is dictated by shear and radiation pressure. The fragment size
distribution is such that at least half of the optical depth is contributed by grains
large enough that their orbits relative to the original body are dominated by shear;
forward shearing is more rapid than backward shearing, so the clump has a sharper rear
edge than front edge, accounting for the different number of positive and negative
gradients measured for $f_{\rm min} < 0.5$. The fragments also comprise small grains
whose dynamics are dominated by radiation pressure, which form a ``tail'' much like a
comet's and account for the egress where $f_{\rm min} > 0.5$. We leave the development of
a quantitative study of this scenario for the future, noting that tests of such a model
would require photometry at multiple wavelengths.

A look at specific dimming events shows that such a simple scenario will face challenges,
as shown by the first set of events in 2006 (Figure \ref{fig:waspkeltzoom}); following
the $f \approx 0.5$ dip there are two more nights of data that have higher average
fluxes, but both nights actually have negative gradients. This evolution does not
invalidate the above analysis, but shows that the temporal evolution is complex, and at
any given time multiple clumps, which may or may not be related, could be occulting the
star.

\begin{figure}
  \begin{center}
    \hspace{-0.5cm} \includegraphics[width=0.7\textwidth]{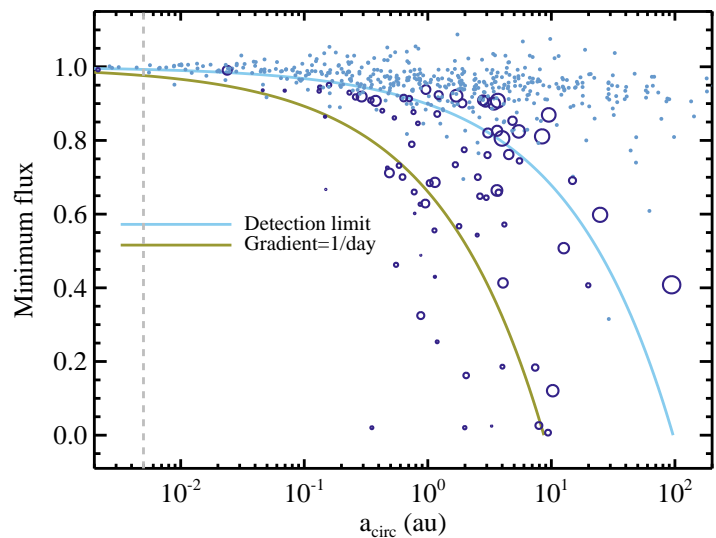}
    \caption{Semi-major axes estimated from the light curve gradients in Figure
      \ref{fig:grad}. Open circles have gradients significantly different from zero, and
      a symbol size proportional to the inverse of the gradient uncertainty. Dots are
      consistent with zero slope. The stellar radius (grey dashed line) has been
      estimated as Solar. Because gradients may have lower minimum fluxes they can move
      down along lines parallel to the solid lines. Points near or above the blue line
      are consistent with zero light curve gradient.}\label{fig:sma}
  \end{center}
\end{figure}

In Figure \ref{fig:sma} we have converted the gradients into semi-major axes using
equation (\ref{eq:acl}). The dashed line shows the stellar radius, estimated to be
approximately Solar based on an age of 25Myr \cite{2013AstL...39..776P} and an effective
temperature of 5350K \cite{2014A&A...563A.139P} using Siess et
al. \cite{2000A&A...358..593S} isochrones. This Figure also includes solid lines of
constant gradient, computed using equation (\ref{eq:acl}) and $f_{\rm min} = 1-\tau$.

The blue line shows the semi-major axis implied by the scatter in gradients near
$f_{\rm min}=1$ in Figure \ref{fig:grad}. Above this line the gradients can be considered
consistent with zero, and the inferred semi-major axes largely meaningless. A related
point is that because the points near $f_{\rm min}=1$ have very small $\tau$, the derived
velocities and semi-major axes can be rather extreme and should be disregarded. An
additional issue is highlighted by the two large circles near $f_{\rm min}=0.5$ and
between $a_{\rm circ}=30$ to 100 au, which are the first two nights of 2006 observations
shown in Figure \ref{fig:waspkeltzoom}. The first is at the time of minimum flux, and the
other also appears to be at a turning point, so the assumption behind equation
(\ref{eq:acl}) that $u=0$ (i.e. the cloud edge is passing the stellar disk center) is
clearly incorrect. These points should therefore be associated not with clumps at 30 to
100au, but with clumps that lie somewhere interior.

The green line shows the semi-major axis implied by a gradient of 1/day. As the lowest
flux measured on a given night is not necessarily the minimum flux for that dimming
event, the inferred semi-major axis could be greater; in this case points move downward
parallel to this and similar lines. Figure \ref{fig:sma} suggests that the clumps orbit
with projected velocities that are consistent with circular orbits at or beyond about
1au. The conclusion from the IR excess, and possibly the periodicity analysis, is that
they may lie closer, near 0.3-0.7au. Reconciling these differences requires that the
clumps are on eccentric obits. For a semi-major axis of 0.3au the minimum clump
eccentricity (i.e. all clumps transit at apocenter) is about 0.6. Such an orientation is
of course highly unlikely, so either the semi-major axis is larger, or the true
eccentricities need to be higher. Perhaps coincidentally, an eccentricity of 0.6 yields a
pericenter velocity of 120km s$^{-1}$, similar to the velocity shifts in Sodium
absorption lines seen towards RZ~Psc, which are discussed further in section
\ref{s:disk}\ref{ss:orig} (noting however that the pericenter velocity is tangential, and
the absorption lines show projected radial velocity \cite{2013Ap.....56..453P}).

\subsection{Summary of light curve analysis}\label{ss:wheresum}

The aim of this section was to estimate the location of the bodies that cause the dimming
events towards RZ~Psc. Using ten years of WASP and KELT-North data, we investigated both
the periodicity of possible repeat events, and the light curve gradients. While we found
evidence for a 60-70 day period, this signal is weak and not statistically
significant. This period would place the clumps near 0.3au, similar to the distance
inferred for the dust seen as an IR excess (though this location is also uncertain). The
gradient analysis places loose constraints on the clump orbits ($<$10au), so is
consistent with a scenario where the clumps have semi-major axes near 0.3au. Thus, based
on the light curve analysis there is no reason to disfavour the model proposed by de Wit
et al. \cite{2013A&A...553L...1D}, that the dimming events are associated with clumps
being created by planetesimal collisions within an asteroid belt analogue.

The primary uncertainty lies with the disk evolutionary state. If a significant gas
reservoir remains, a turbulent inner rim scenario similar to that proposed for UXors
might produce a similar light curve. Both the lack of accretion, and a likely age beyond
which gas-rich disks are typically seen, may argue against this scenario, though the disk
may be in transition to the debris phase and retain some primordial gas. We revisit the
disk status from the perspective of the flux distribution in section
\ref{s:disk}\ref{ss:evol}.

An additional result from the gradient analysis concerns the structure of individual
clumps. The non-uniform gradient distribution is qualitatively consistent with
post-collisional asteroidal fragments being dispersed by a combination of shearing and
radiation pressure. Assuming small dust well-coupled to gas in a hydrodynamic turbulence
scenario, a uniform gradient distribution seems more likely, so we interpret the
gradients as providing circumstantial evidence for the planetesimal fragment scenario.

\section{Disk structure and evolutionary state}\label{s:disk}

Based on ten years of relatively high-cadence photometry, RZ~Psc is regularly occulted by
what are almost certainly star-sized clumps of dust. These clumps can be optically thick,
and previous measurements of colour variations show that at least some of the dust must
be small (e.g. \cite{2003ARep...47..580S}), which may be supported by the light curve
gradient statistics. The previous interpretation of this system was that the clumps are
the fragments arising from planetesimal collisions within an asteroid belt analogue that
is also detected in the mid-IR. While the results from the previous section are
consistent with this scenario, the evidence is at best circumstantial as they do not rule
out the alternative of an UXor-like hydrodynamic inner rim scenario.

We now address several open questions, each taking a slightly wider view. Primary among
these is whether the dimming events and the IR excess are caused by the same dust, as
suggested by de Wit et al. \cite{2013A&A...553L...1D}. Two further aspects are then the
implications for the origin of the clumps and the evolutionary status of the disk in
which they reside. We finish by considering the proposed scenarios for dippers and UXors,
and why RZ~Psc appears to be a rare object that lies between these classes.

\subsection{Does the occulting dust account for the IR excess?}\label{ss:ir}

One of the reasons that RZ~Psc is worthy of detailed study is that circumstellar dust is
inferred from both the dimming events and the IR excess. Different properties of the dust
grains, and the larger structure in which they reside, are revealed by each method; the
dimming events yield information on dust ``clumpiness'' on a star-sized scale, while the
IR excess provides evidence for a disk that captures $\sim$7\% of the starlight, and thus
a measure of the total surface area of dust. The proposed interpretation is that the
clumps orbit within an asteroid belt, and the dimming events therefore provide some
information on the size distribution and collisional evolution within the belt
\cite{2013A&A...553L...1D}. This expectation relies on co-location of the clumps and the
belt, for which circumstantial evidence is provided by the light curve gradients and
perhaps the periodicity analysis (see also \cite{2010A&A...524A...8G}).

The fraction of starlight intercepted by the dust is a variable common to both the
dimming events and the IR excess. For the former we use the average extinction $\bar{E}$,
which is simply taken from the normalised light curve, as 1 minus the average flux,
yielding 0.05 (the light curve median is 0.995). This estimate assumes that all dimming
events are independent, and the value would be smaller if not since the dust in some
clumps may be being counted two or more times. The lack of strong evidence for
periodicity suggests that multiple counting is not a serious issue however. Another issue
is that the star could be reddened, and therefore that the normalised light curve has
already had some constant level of extinction removed. Based on photospheric colours this
unseen extinction is probably small, in the range of zero to a few percent
\cite{2000ARep...44..611K}.

If we assume that this average extinction applies over a uniform sphere around the star,
and that the dust has a low albedo (i.e. the dimming events are dominated by dust
absorption, not scattering of light out of our line of sight), then the IR fractional
luminosity is equal to the average extinction. That is, both are equal to the fraction of
starlight intercepted by the dust.

To explore possible geometries we use a simple relation between fractional luminosity
$L_{\rm disk}/L_\star$, (uniform) geometric optical depth $\tau$, and the disk opening
angle $\theta$ \cite{2014MNRAS.438.3299K}
\begin{equation}
  L_{\rm disk}/L_\star = \tau sin(\theta/2) \, ,
\end{equation}
which says that the fractional luminosity is the optical depth of the dust multiplied by
the fraction of the sky covered as seen from the star. The dust belt must therefore have
an opening angle of at least 8$^\circ$ to capture 7\% of the starlight. However, for this
minimal estimate the dust is optically thick, yet RZ~Psc is not seen to be reddened. If
we instead require $\tau \sim \bar{E}$ then as stated in the previous paragraph the dust
distribution must instead be near isotropic. Given the ubiquity of disk-like structures
around young stars, such a spherical distribution seems physically unlikely. In addition,
the increased polarisation during deep dimming events argues against a spherical
distribution.

Thus, the picture of RZ~Psc as a star seen \emph{through} a disk, where the clumps
account for all of the dust and sample some representative part of an asteroid belt
(e.g. the midplane), is untenable because that belt would cause much more reddening than
is observed. These characteristics distinguish RZ~Psc from heavily reddened objects,
where the dimming events could be sampling a more representative section of the disk
(e.g. \cite{2015MNRAS.451...26S}). This issue can be avoided by invoking a spherical
distribution of material, but the problem then shifts to whether such a distribution is
physically plausible.

A more likely alternative, which we favour, is that most of the dust does not lie on
orbits that pass in front of the star, and the occultations are caused by a small
fraction of objects that have higher vertical locations (or greater orbital inclinations)
than average. In this case the component that causes most of the IR excess may or may not
be clumpy, and could be radially optically thick (i.e. the opening angle could be as
small as 8$^\circ$). This picture unfortunately loses any strong connection between the
occulting clumps and the IR excess, essentially adding a free parameter that is the
fraction of material that is ``kicked'' or resides above the disk, but seems to be the
simplest and most probable scenario. Dullemond et al. \cite{2003ApJ...594L..47D} used
essentially the same argument for UXors, so our picture is therefore inevitably similar
to dust occultation models proposed for UXors
(e.g. \cite{1997ApJ...491..885N,2000A&A...364..633N,2003ApJ...594L..47D}).  As the disk
is probably radially optically thick with a scale height similar to gas-rich
protoplanetary disks, it could be that the scenario for RZ~Psc is in fact the same as
proposed for UXors. In this case, the IR excess would originate from the inner edge of a
more extended disk, which is not detected at longer wavelengths for reasons discussed
below.

We therefore conclude that while there is almost certainly some connection between the
dimming events and the IR excess, it is at best indirect; we are not viewing RZ~Psc
through a representative part of an asteroid belt analogue. As with other UXors, a clear
prediction is that the disk is not seen edge-on, but at an intermediate inclination.

\subsection{Origin of the occulting structures}\label{ss:orig}

One of the distinguishing characteristics for RZ~Psc is the relatively short duration of
the dimming events $t_{\rm dim}$, which are a few days compared to a few weeks for other
UXors (e.g. \cite{1999AJ....118.1043H,2010A&A...511L...9C}). If we assume near-circular
orbits at speed $v_{\rm kep}$, $t_{\rm dim}=2(R_{\rm cl}+R_\star)/v_{\rm kep}$, where the
clump has radius $R_{\rm cl}$. Solving for the clump radius yields the relation
(e.g. \cite{2016ApJ...816...69A,2016MNRAS.457.3988B}):
\begin{equation}
  R_{\rm cl} \approx 1.85 t_{\rm dim} \, \left( \frac{M_\star}{M_\odot} \frac{1 {\rm au}}{a} \right)^{1/2} - R_\star \, ,
\end{equation}
where here $R_{\rm cl}$ and $R_\star$ are in units of $R_\odot$ and $t_{\rm dip}$ is in
days. This equation says that dimming events of a given duration can in general be caused
by larger clumps that orbit close to the star, or smaller clumps that orbit farther out.


\begin{figure}
  \begin{center}
    \hspace{-0.5cm} \includegraphics[width=0.7\textwidth]{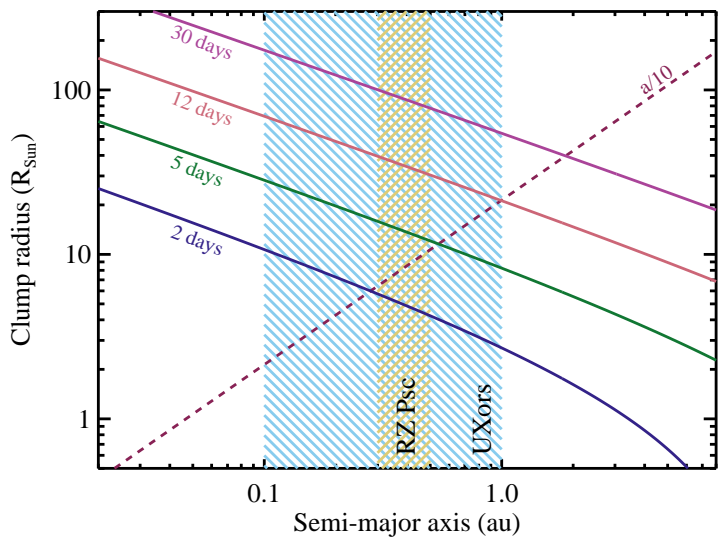}
    \caption{Clump properties assuming circular orbits for a range of dimming event
      durations (as labelled). The brown region marks the approximate location of the
      RZ~Psc asteroid belt (or the inner edge of a more extended disk). The blue shaded
      region shows the range of Herbig Ae inner disk edge radii. The dashed line shows
      where a clump has an azimuthal extent similar to the scale height of a typical
      gas-rich disk.}\label{fig:rcla}
  \end{center}
\end{figure}

Figure \ref{fig:rcla} shows the radius that clumps must have to cause dimming events of
different durations as a function of semi-major axis. For clumps much larger than the
star the stellar radius is unimportant, and the stellar mass dependence is relatively
weak, so this plot can be applied to RZ~Psc and UXors. The radius is of course the
sky-projected size of a clump along the orbit, so whether this scale also applies
vertically and radially depends on the specific scenario. This plot assumes circular
orbits, and that the star has Solar radius and mass. The approximate locations of the
RZ~Psc belt (or the inner edge of a more extended disk), and the range of inner edge
radii for Herbig Ae stars \cite{2007prpl.conf..539M}, are shown by the hatched regions.

The dashed line shows where clumps extend a tenth of the semi-major axis - approximately
the scale height for a gas-rich disk. If the variability of both UXors and RZ~Psc
originates from dust structures arising from hydrodynamic turbulence at the disk inner
edge, and these structures are related to the disk scale height, then a wide range of
dimming event times is expected. These times should correlate strongly with the inner
edge location, and because this location is set by sublimation
\cite{2007prpl.conf..539M}, should correlate with the luminosity of the star. We did not
find evidence for such a correlation in time-variability studies of UXors
(e.g. \cite{1991A&AS...89..319B}), suggesting that the clump radii do not vary strongly
with the inner edge location. In any case, we have already noted that RZ~Psc has shorter
dimming events than ``typical'' UXors, so while RZ~Psc has dust at a radius that falls
within the range of UXors, the occulting clumps are inferred to be several times smaller.

A further difference between RZ~Psc and UXors is the origin of the dust location. For
Herbig Ae/Be stars (and by extension, UXors) the dust inner radius is set by
sublimation. However, the dust around RZ~Psc is roughly 500K, so much cooler than the
$\sim$1500K sublimation temperature. That is, if the origin of RZ~Psc's variability is
interpreted as similar to other UXors and originates in a gas-rich disk, it must host a
transition disk rather than a ``full'' primordial disk. Therefore, while these
comparisons show RZ~Psc to be unusual compared to typical UXors, they do not argue
strongly for or against a specific scenario.

A final aspect to discuss regarding the origin of the clumps, and their relation to the
IR excess, is the transient absorption features. These are seen towards UXors, but also
seen towards some main-sequence A stars
(e.g. \cite{1987A&A...185..267F,2013PASP..125..759W,2014A&A...561L..10K}), so are not
exclusive to stars that host gas-rich disks. For A-type stars these features are
generally interpreted as sun-grazing ``exocomets'', and the same may apply to UXors and
RZ~Psc. A potential issue with this interpretation is that the absorption lines towards
RZ~Psc are so far blue-shifted, and may instead originate in an outflow
\cite{2013Ap.....56..453P}. However, blue-shifting could also occur if evaporation mostly
occurs near periastron passage, which might be expected if the bodies originate in an
asteroid belt and are thus more refractory than the exocomets seen towards other stars
(i.e. an Asteroid in the Solar System would need to pass very close to the Sun to have a
tail). Such a scenario is also consistent with the conclusion of section \ref{s:where}
that the occulting clumps could be on eccentric orbits with high velocities at
periastron. Models of such low-periastron asteroids or comets invariably require a
perturbing planet (e.g. \cite{1990A&A...236..202B,1996Icar..120..358B}). Assuming the
same 4:1 mean-motion resonance picture of Beust \& Morbidelli \cite{1996Icar..120..358B}
implies a planet about 2.5 times more distant than the source. That is, if the asteroid
belt is at 0.3au, the planet is near 0.75au. It may be possible that this planet is
inclined and causes the asteroid belt to precess, thus causing the 12.4 year modulation
of the stellar flux seen by \cite{2013A&A...553L...1D}.

Overall, the origin of the clumps remains unclear, and raises many further questions. A
distinguishing feature of RZ~Psc's light curve is that most of the time the star is near
the quiescent level; if the dimming events are related to variable structure at the inner
rim of a gas-rich disk, then why are the events so rare? Is the geometry so finely tuned
that only the most extreme fluctuations are visible? In the asteroid belt scenario the
details are equally unclear; if the clumps are recently disrupted planetesimals, how do
they appear above the bulk of the disk when collisions are most likely to occur at the
mid plane? Are there two populations of objects, one that forms the main disk, and
another that causes the dimming events? Could such a scenario be reconciled with the
transient absorption features? Answers to these questions will require further study, and
will almost certainly require new observations.

\subsection{Disk evolutionary state}\label{ss:evol}

Many young stars with gas-rich disks show broadly similar variability (UXors, AA Tau
analogues, dippers). With sporadic and deep photometric minima, and transient absorption
lines, RZ~Psc is most similar to UXors and has often been associated with this class,
albeit as an unusual member (e.g. \cite{2010A&A...524A...8G}). Initial distinctions were
made based on the late spectral type (K0V), the relatively short dimming events, and a
lack of near IR excess and accretion signatures. The more recent findings that the
stellar age is probably several tens of Myr, and that the IR excess is suggestive of an
Asteroid belt analogue, further distinguish RZ~Psc as a potentially remarkable object
where the early gas-poor stages of main-sequence debris disk evolution can be
studied. There remain similarities between RZ~Psc and UXors. Primarily, we concluded that
the clumps that cause the dimming events are most likely seen when they are well above
the densest regions of a disk, consistent with the turbulent inner rim scenario proposed
by Dullemond et al. \cite{2003ApJ...594L..47D}.

\begin{figure}
  \begin{center}
    \hspace{-0.5cm} \includegraphics[width=0.7\textwidth]{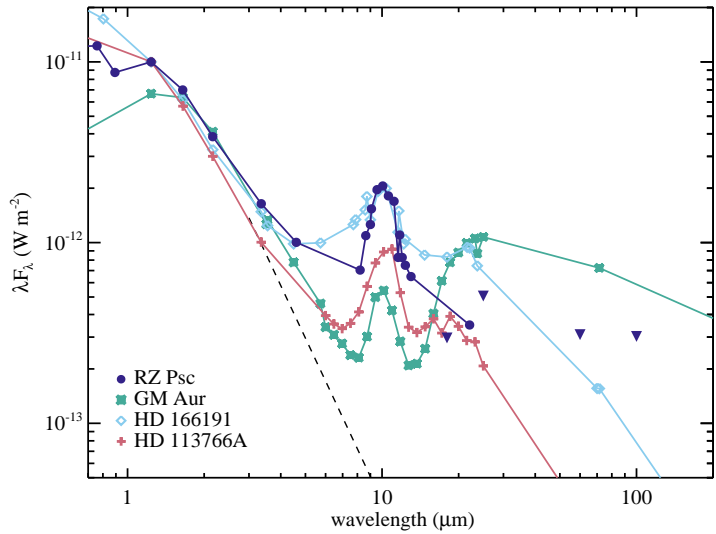}
    \caption{The SED of RZ~Psc in comparison with other young disk-hosting stars. All
      SEDs are normalised to a common flux density at $H$ band. The triangles are AKARI
      and IRAS upper limits for RZ~Psc. Measurements we made at different times, so the
      apparent discrepancy between the WISE detection at 22$\mu$m and the AKARI upper
      limit at 18$\mu$m is an indicator of IR variability (see section
      \ref{s:data}\ref{ss:irvar}).}\label{fig:spcomp}
  \end{center}
\end{figure}

To consider RZ~Psc within the context of other young disk-hosting systems, Figure
\ref{fig:spcomp} shows the spectral energy distribution (SED) of RZ~Psc, and several
other systems that could be considered to be at a similar evolutionary stage (a similar
plot appeared in \cite{2014MNRAS.438.3299K}). GM Aur hosts a transition disk
\cite{2005ApJ...630L.185C}, the status of HD 166191's disk is ambiguous and may lie
somewhere between the transition and debris phase
\cite{2013ApJ...777...78S,2014MNRAS.438.3299K}, and HD 113766A is generally considered to
host a bright warm debris disk (based on a 10-16Myr age and a lack of gas
\cite{2006ApJS..166..351C}, but based on similarities with HD~166191 was also noted as
potentially ambiguous \cite{2014MNRAS.438.3299K}).

In contrast to UXors, there is no obvious reason for dust around RZ~Psc to lie near 0.3
au, as the dust sublimation distance is much closer. For RZ~Psc to host a gas-rich disk
it would therefore probably need to be a transition object that has as-yet undetected
far-IR emission from the outer disk. As illustrated by Figure \ref{fig:spcomp} the limits
set by IRAS are not particularly stringent, but in comparison to GM Aur the SED beyond
20$\mu$m is a factor of two lower, despite the mid-IR SED being a factor of two
brighter. Transition disks have a wide variety of spectra however
(e.g. \cite{2011ApJ...728...49E,2014MNRAS.443.1587R}, so the main conclusion from this
comparison is that if an outer disk exists, it is not very bright. This far-IR deficit is
a signature of self-shadowing, which is of course a key characteristic of UXors. If such
a picture were true for RZ~Psc, a prediction is that the outer disk may still be
detectable at millimeter wavelengths (roughly mJy levels), whereas an extrapolation based
on the asteroid belt scenario would not (roughly $\mu$Jy). Specifically, self-shadowing
can be caused by settling of dust towards the disk mid-plane, which can occur with little
loss of vertical optical depth \cite{2004A&A...421.1075D}. Thus the far-IR flux can be
much lower than for a typical disk, while the mm-wave flux is not.

In Figure \ref{fig:spcomp}, RZ~Psc looks more similar to HD~166191 and HD~113766A, thus
falling in the category of systems whose interpretation in terms of disk status is
ambiguous (the high fractional luminosity would also mark RZ~Psc as unusually extreme for
a debris disk). With the aid of mid-IR interferometry, the disk around HD 113766A has
been shown to comprise two components, one at 0.6 and another at 9au
\cite{2013A&A...551A.134O}. This is by no means evidence that RZ~Psc has a similar
structure, but merely reinforces the fact that the SED does not rule out such
possibilities and that the dust need not be confined to a single belt near 0.3au, and
need not be interpreted as an asteroid belt. Similarly, the disk around HD 166191 was
modelled as an optically thick transition disk extending from 1-25au
\cite{2014MNRAS.438.3299K}. Thus, as was concluded above by considering self-shadowing,
it could be that RZ~Psc hosts a disk that extends from 0.3 to a few tens of au.

While the lack of a large near-IR excess for RZ~Psc suggests that the disk is at least in
the transition to a debris disk (i.e. has an inner hole), it does not preclude the
possibility that gas resides in that hole and may still be accreting on to the star. No
emission lines that would provide evidence of accretion have been seen
\cite{2013Ap.....56..453P,2014A&A...563A.139P}, but as a further test we reconsidered the
spectral energy distribution. Specifically, we included photometry from the Galaxy
Evolution Explorer (GALEX \cite{2003SPIE.4854..336M}) to quantify the level of any
ultraviolet (UV) excess, a complementary accretion indicator
(e.g. \cite{1998ApJ...509..802C}). This exercise is hindered somewhat by the possibility
that optical photometry was obtained when RZ~Psc was not near the quiescent level. To
circumvent this issue we used just the Two Micron All-Sky Survey (2MASS
\cite{2003tmc..book.....C}) and GALEX photometry, fitting a PHOENIX atmosphere model
\cite{2005ESASP.576..565B}, finding a best-fit effective temperature of 5485K (assuming
no reddening, or 5600K if some reddening is allowed to improve the fit slightly). These
temperatures are consistent with that derived from a high-resolution spectrum;
$5350 \pm 150$K \cite{2014A&A...563A.139P}. Alternatively, fixing the temperature to the
spectroscopic value yields a mild UV excess, less than a factor of two, that may be
chromospheric. We therefore conclude that there is no evidence for accretion seen as a UV
excess.

Finally, the mid-IR variability discussed in section \ref{s:data}\ref{ss:irvar} provides a
measure of the changing emitting area of the disk around RZ~Psc, and thus potentially
information about the disk structure and status. Figure \ref{fig:wisevar} shows that for
the epochs where contemporaneous 3-22$\mu$m photometry exists there is no evidence of
``seesaw'' variability over 6 months, but that strong conclusions are limited by a lack
of data. Over 5 years the 3-5$\mu$m disk flux varied by about a factor of two, with no
major changes in the behaviour of the optical light curve. However, given that the bulk
of the disk emission probably originates from material that is not occulting the star,
direct links between the optical and IR behaviour is not necessarily expected. Indeed,
towards young stars hosting gas-rich disks, both correlated and uncorrelated optical/IR
variability is seen \cite{2014AJ....147...82C}. Similarly, the IR flux of bright warm
debris disks has been seen to vary strongly while the optical brightness remains constant
\cite{2015ApJ...805...77M}. The main benefit of more intensive IR monitoring would be to
search for mid-IR ``see-saw'' variability, because it would provide evidence that the
disk around RZ~Psc has a significant radial extent.

In summary, the status of the disk surrounding RZ~Psc is unclear. Despite a reasonable
near/mid-IR characterisation, at longer wavelengths the SED is not detected. Comparison
with the infrared spectra of other disks suggests that the disk is at a minimum well
evolved towards the debris phase, and may have reached it already. This possibility, and
the rarity of objects like RZ~Psc, opens the possibility that it is being observed at a
special time with a specific geometry, so may yield better insights than most systems.

\subsection{The rarity of Sun-like dippers/UXors}\label{ss:rarity}

\begin{figure}
  \begin{center}
    \hspace{-0.5cm} \includegraphics[width=0.7\textwidth]{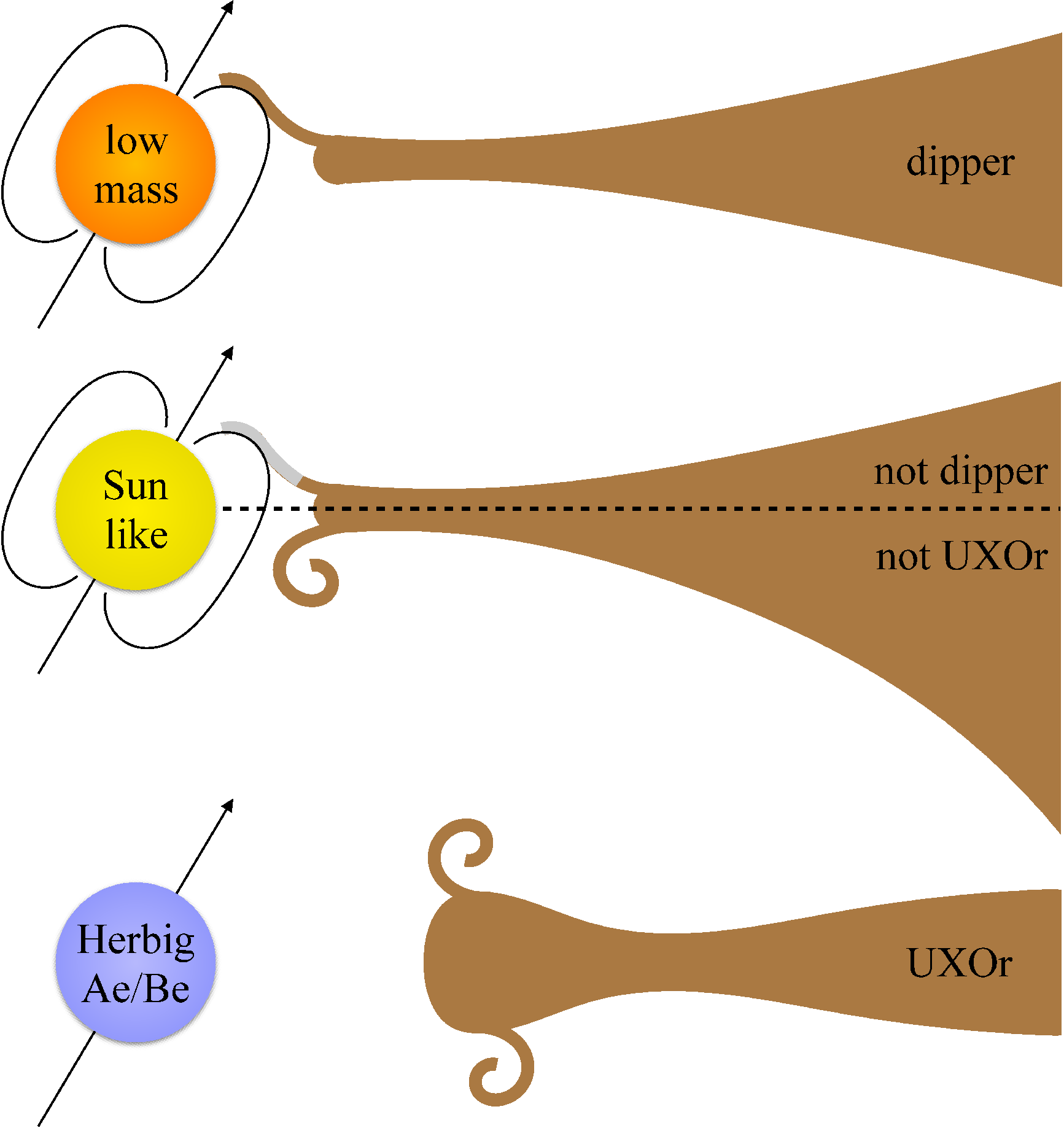}
    \caption{Cartoon showing possible origins of dippers and UXors, and why Sun-like
      stars may only rarely show analogous behaviour. In each case the star, magnetic
      dipole, and rotation axis are shown at the left (the stellar magnetic field is not
      necessarily always tilted with respect to the disk). Possible disk structures
      viewed edge-on to the right. (\emph{top}) low-mass stars (dippers) are occulted by
      co-rotating material that is accreting on to the star, and the dust sublimation
      radius is interior to corotation \cite{2016arXiv160503985B}. (\emph{middle})
      Sun-like stars are rarely seen as dippers or UXors because i) dust sublimates
      outside corotation (represented by the grey accretion column in the upper half
      \cite{2016arXiv160503985B}), or ii) material lifted by turbulence is shadowed by
      the outer disk (spiral in the lower half). (\emph{bottom}) Herbig Ae/Be stars
      (UXors) are occulted by turbulence that appears above self-shadowed disks
      \cite{2003ApJ...594L..47D}.}\label{fig:cartoon}
  \end{center}
\end{figure}

We finish by briefly discussing RZ~Psc in the context of the general classes of dippers
and UXors, considering why neither of these classes includes many young Sun-like stars
like RZ~Psc. To aid this discussion, Figure \ref{fig:cartoon} shows simplified cartoons
of the proposed scenarios for dippers and UXors, and possible reasons for a lack of
significant dust-related dimming events towards Sun-like stars (see
\cite{2016arXiv160503985B} for similar figures for dippers).

From the perspective of dippers, which so far have spectral types later than K5, Bodman
et al.  \cite{2016arXiv160503985B} explain their tendency to be low-mass stars as a
consequence of the relation between the magnetospheric truncation, corotation, and
sublimation radii. The periodicity of dippers suggests that the occulting material is
near the corotation radius, and is therefore near the base of any accretion columns. For
low-mass stars the dust temperature at this distance is cool enough that the columns
contain significant dust mass and hence the dipping phenomenon is seen (top panel of
Figure \ref{fig:cartoon}. For earlier type stars the sublimation radius is outside the
corotation radius, so any dust in the accretion columns has sublimated and no dipping is
seen (grey accretion column in the middle panel of Figure \ref{fig:cartoon}).

An as-yet unexplored corollary of this scenario is that the outer disks in dipper systems
must not be more flared than the height of the accretion columns (as seen from the
star). This could be because disks around low mass stars are simply less flared in
general (e.g. \cite{2010ApJ...720.1668S}), because disks in dipper systems tend to be
more evolved than average \cite{2016ApJ...816...69A}, or because there is less relation
between the outer disk geometry, the inner disk, and accretion columns than would be
naively expected \cite{2016MNRAS.462L.101A}. Of course, if disks around Sun-like stars
are sufficiently flared that the inner regions are not visible, then whether the
accretion streams are transparent or not is moot.

From the perspective of UXors, a possible explanation for their tendency to be late B and
A-type stars is that the specifics of self-shadowing are different for Sun-like stars
\cite{2003ApJ...594L..47D}. There is little reason to believe that self-shadowing does
not happen for young Sun-like stars \cite{1987ApJ...323..714K,2004A&A...421.1075D}, so it
seems either that self-shadowing is simply rarer, or that the nature is different in a
way that affects whether the inner disk can occult the star (i.e. shadowing is by a
larger portion of the inner disk regions rather than by a puffed-up inner rim,
\cite{2004A&A...421.1075D,2007prpl.conf..555D}).

Thus, it seems that the rarity of Sun-like stars among dipper and UXor populations may be
understood as a result of the scenarios for both. Accretion columns are optically thin
because the dust has sublimated, and any turbulence that rises above the inner disk may
not be seen as it is hidden behind a flaring outer disk.

In this context the rarity of RZ~Psc-like objects can be explained in two ways. The first
simply sidesteps the above discussion by interpreting the disk as a gas-poor asteroid
belt analogue. Examples of such bright disks at a few au are rare
\cite{2013MNRAS.433.2334K}, and only a subset of these will be oriented such that dimming
events are seen above the disk mid plane (recalling that the occulting bodies are at tens
to hundreds of stellar radii). The second explanation relies on RZ~Psc's age and SED,
which suggest that it hosts a gas-rich transition disk that is well settled (i.e. not
significantly flared), and hence turbulence above the disk inner edge is visible. The
non-detection of RZ~Psc in the far-IR (Figure \ref{fig:spcomp}) might also be the result
of such settling, and suggests that the brightness at millimeter wavelengths might be
brighter than expected given the mid/far-IR brightness
(e.g. \cite{2004A&A...421.1075D,2007prpl.conf..555D}). The rarity is again explained by
an unlikely geometry, and perhaps that the period during which the inner disk can be seen
above the outer disk as it settles is relatively short.

\section{Summary and Conclusions}\label{s:conc}

Long considered a member of the UXor class of variables, RZ~Psc is almost completely
occulted by dust for several days, multiple times during each observing season. The
``typical'' UXor (which is generally a Herbig Ae object) shows day to month long dimming
events that are thought to be caused by hydrodynamic turbulence above the disk inner rim,
and where the outer disk is self-shadowed \cite{2003ApJ...594L..47D}. Various anomalous
characteristics distinguish RZ~Psc from other UXors; the possible age of a few tens of
Myr, the K0V spectral type, the few-day long dimming events, and the location of the
IR-excess emitting dust well beyond the sublimation radius. These characteristics have
been used by de Wit et al. \cite{2013A&A...553L...1D} to suggest that RZ~Psc hosts a
gas-poor asteroid belt analogue at 0.4-0.7au and that the dust clumps that occult the
star are the dispersed fragments produced in destructive planetesimal collisions.

To take a critical look at this intriguing scenario, we have presented and analysed ten
years of WASP and KELT-North photometric monitoring of RZ~Psc. We found circumstantial
evidence that some dimming events repeat and have a semi-major axis consistent with that
inferred from the IR excess, but the signal is not significant (1-2$\sigma$). The light
curve gradients are consistent with this picture, but the constraints are poor. The
statistics of the light curve gradients suggest that a typical dimming event has an
egress rate that is initially faster, and then slower, than ingress. While this evolution
seems qualitatively consistent with the structure expected from a planetesimal collision,
quantitative models are needed.

By considering the joint constraints allowed by the light curve and the IR excess, we
find that the objects causing the dimming events are unlikely to be representative of the
structure causing the IR excess. The two can only be reconciled if the IR excess
originates in a spherical shell of clumpy but on average optically thin dust, a scenario
disfavoured by the increased polarisation during deep dimming events. Assuming a
disk-like structure, the belt is almost certainly optically thick with an opening angle
of a few tens of degrees, with the system viewed at an inclination that allows clumps
residing above this belt to pass in front of the star. While such a geometry is possible
for both UXor-like and asteroid belt scenarios, the relatively cool temperature of the
dust around RZ~Psc means that if it hosts a gas-rich disk it must be a transition
object. Indeed, comparison of RZ~Psc's spectrum with other objects suggest that it is
similar to objects whose status is ambiguous, and could be either debris or transition
disks. The low far-IR disk luminosity could arise if the outer disk is shadowed (as
suggested for UXors), or simply because there is no outer disk. The spectrum is poorly
sampled and would benefit from millimeter photometry, specifically to test whether there
is a settled outer disk.

Overall, we conclude that the status of RZ~Psc's disk is uncertain, and therefore that so
is the origin of the clumps. The lack of near-IR excess shows that the disk is beyond the
primordial phase, but could be in the final throes of dispersal and the occulting
structures a related phenomenon. Several specific observations would help: i) mid-IR
spectral monitoring would allow comparisons with transition disk systems that show
``see-saw'' variability, ii) continuous photometry (ideally with multiple colours) would
yield the detailed shape of individual dimming events and in some cases the distribution
of dust size across the clump.

As a young Sun-like star showing disk-related stellar variability, RZ~Psc is a
rarity. The reason may be twofold; i) the sublimation radius is greater than for low-mass
stars so any accretion streams are transparent, and/or ii) in contrast to more massive
stars, turbulence above the inner rim may be shadowed by the outer disk. While the
asteroid belt scenario avoids the need to consider primordial disk structure, in the
context of such models RZ~Psc-like variability could be explained by the evolved state of
the disk, which may have settled enough that the inner rim is visible.

\ack{}

\textbf{Data and code:} Materials used in the preparation of this contribution can be
found online at \href{https://github.com/drgmk/rzpsc}{https://github.com/drgmk/rzpsc}.

\textbf{Competing interests:} We declare that we have no competing interests.

\textbf{Contributions:} GMK initiated the project, collated data, did most of the
analysis, and wrote the paper. MAK contributed to the time-series analysis and conceived
the iterative event finding method. KGS provided independent SED models to test for a UV
excess. As members of the KELT collaboration, JP, JER, RJS, and KGS acquired the
time-series photometry. MCW was a co-I on the proposal to obtain the VISIR data. All
co-authors provided input on the style and content of the manuscript.

\textbf{Acknowledgments:} We thank Joachim G\"urtler for sharing photometry from the
Sonneberg and Harvard Plates in a palatable form, AAVSO observers for monitoring RZ~Psc,
and Simon Hodgkin and Jim Pringle for useful discussions. This paper makes use of data
from the DR1 of the WASP data \cite{2010A&A...520L..10B} as provided by the WASP
consortium, and the computing and storage facilities at the CERIT Scientific Cloud,
reg. no. CZ.1.05/3.2.00/08.0144 which is operated by Masaryk University, Czech Republic.

\textbf{Funding:} GMK is supported by the Royal Society as a Royal Society University
Research Fellow. JER is funded as a Future Faculty Leaders Fellow at the
Harvard-Smithsonian Center for Astrophysics. MCW acknowledges support from the European
Union through ERC grant number 279973. Early work on KELT-North was supported by NASA
Grant NNG04GO70G.



\end{document}